\documentstyle[12pt,aps,epsf]{revtex}
\def\cond {\langle \varphi \rangle}
\newcommand{\be}{\begin{equation}}
\newcommand{\ee}{\end{equation}}
\newcommand{\beq}{\begin{eqnarray}}
\newcommand{\eeq}{\end{eqnarray}}
\begin{document}
\tighten{

\title{{Particle Physics Aspects of Modern Cosmology}
\footnote{Invited lectures at the 15th Symposium on Theoretical Physics, Seoul National University, August 22 - 28 1996; to be
published in {\it Field Theoretical Methods in Fundamental Physics}, ed. by Choonkyu Lee (Mineumsa Co. Ltd., Seoul, 1997).}}
\author{}
\author{}
\author{Robert H. BRANDENBERGER\\ 
Department of Physics, Brown University\\
Providence, RI 02912, USA}
\author{} 
\author{}
\maketitle

\bigskip
\begin{abstract}
\noindent
Modern cosmology has created a tight link between particle physics / field theory and a wealth of new observational data on the structure of the Universe. These lecture notes focus on some of the most important aspects concerning the connection between theory and observations. The lectures begin with an overview of some recent progress and problems in inflationary cosmology. In particular, a pedagogical discussion of the theory of reheating is presented. The second topic is a survey of the theory of cosmological perturbations, the cornerstone of modern cosmology. The focus is on the gauge-invariant classical and quantum theory of fluctuations. The third topic concerns the role of topological defects in cosmology. Reviews of the cosmic string theory of galaxy formation and of defect-mediated GUT and electroweak baryogenesis are given. 
\end{abstract}

\vskip 0.8cm \noindent {BROWN-HET-1067\\
astro-ph/9701xxx}
\newpage
 
\section{Introduction and Outline}

Most aspects of high energy physics beyond the standard model can only be tested by going to energies far greater than those which present accelerators can provide. Fortunately, the marriage between particle physics and cosmology has provided a way to ``experimentally" test the new theories of fundamental forces.

The key realization, discovered both in the context of the inflationary Universe scenario$^{\cite{Guth}}$ and of topological defects models$^{\cite{ZelVil}}$ is that physics of the very early Universe may explain the origin of structure in the Universe. It now appears that a rich set of data concerning the nonrandom distribution of matter on a wide range of cosmological scales, and on the anisotropies in the cosmic microwave background (CMB), may potentially be explained by high energy physics. In addition, studying the consequences of particle physics models in the context of cosmology may lead to severe constraints on new microscopic theories. Finally, particle physics and field theory may provide explanations of some deep cosmological puzzles, e.g. why the Universe at the present time appears so homogeneous, so close to being spatially flat, and why it contains the observed small net baryon to entropy ratio.

In these lectures, I focus on three important aspects of modern cosmology. The first concerns some fundamental problems of inflationary cosmology. In particular, some recent progress in the understanding of ``reheating" in inflation will be reviewed.

The second topic is the classical and quantum theory of cosmological perturbations, the main tool of modern cosmology. A general relativistic and quantum mechanical analysis of the generation and evolution of linearized fluctuations is essential in order to be able to accurately calculate the amplitude of density perturbations and CMB anisotropies.

As a third topic, I discuss the role of topological defects in baryogenesis and as possible sees for cosmological structure formation. 

The specific outline is as follows:
\begin{enumerate}
\item{} {\bf Introduction and Outline}
\item{} {\bf Lecture 1: Inflationary Universe: Progress and Problems}
\\{2.A} Problems of Standard Cosmology
\\{2.B} Inflationary Universe Scenario
\\{2.C} Problems of Inflation
\\{2.D} Inflation and Nonsingular Cosmology
\\{2.E} Reheating in Inflationary Cosmology
\\{2.F} Summary
\item{} {\bf Lecture 2: Classical and Quantum Theory of Cosmological Perturbations}
\\{3.A} Basic Issues
\\{3.B} Newtonian Theory
\\{3.C} Relativistic Theory: Classical Analysis
\\{3.D} Relativistic Theory: Quantum Analysis
\\{3.E} Summary
\item{} {\bf Lecture 3: Topological Defects, Structure Formation and Baryogenesis}
\\{4.A} Quantifying Data on Large-Scale Structure
\\{4.B} Topological Defects
\\{4.C} Formation of Defects in Cosmological Phase Transitions
\\{4.D} Evolution of Strings and Scaling
\\{4.E} Cosmic Strings and Structure Formation
\\{4.F} Specific Predictions
\\{4.G} Principles of Baryogenesis
\\{4.H} GUT Baryogenesis and Topological Defects
\\{4.I} Electroweak Baryogenesis and Topological Defects
\\{4.J} Summary
\end{enumerate}

Unless otherwise specified, units in which $\hbar = c = k_B = 1$ will be used. Distances are expressed in Mpc (1pc $\simeq$ 3.06 light years). Following the usual convention, $h$ indicates the expansion rate of the Universe in units of $100$ km s$^{-1}$ Mpc$^{-1}$, $\Omega = \rho / \rho_c$ is the ratio of the energy density $\rho$ to the critical density $\rho_c$ (the density which yields a spatially flat Universe), $G$ is Newton's constant and $m_{pl}$ is the Planck mass.
 
\section{Inflationary Universe: Progress and Problems}

The hypothesis that the Universe underwent a period of exponential expansion at very early times has become the most popular theory of the early Universe. Not only does it solve some of the problems of standard big bang cosmology, but it also provides a causal theory for the origin of inhomogeneities in the Universe which is predictive and in reasonable agreement with current observational results. Nevertheless, there are several problems of principle which merit further study.

\subsection{Problems of Standard Cosmology}

The standard big bang cosmology rests on three theoretical pillars: the
cosmological principle, Einstein's general theory of relativity and a perfect
fluid description of matter.

The cosmological principle states that on large distance scales the
Universe is homogeneous. This implies that the metric of space-time can be written in Friedmann-Robertson-Walker (FRW) form:
\be
 ds^2 = a(t)^2 \, \left[ {dr^2\over{1-kr^2}} + r^2 (d \vartheta^2 + \sin^2
\vartheta d\varphi^2) \right] \, , 
\ee
where the constant $k$ determines the topology of the spatial sections. In the following, we shall usually set $k = 0$, i.e. consider a spatially closed Universe. In this case, we can without loss of generality take the scale factor $a(t)$ to be equal to $1$ at the present time $t_0$, i.e. $a(t_0) = 1$. The coordinates $r, \vartheta$ and $\varphi$ are comoving spherical coordinates. World lines with constant comoving coordinates are geodesics corresponding to particles at rest. If the Universe is expanding, i.e. $a(t)$ is increasing, then the physical distance $\Delta x_p(t)$ between two points at rest with fixed comoving distance $\Delta x_c$ grows:
\be
\Delta x_p = a(t) \Delta x_c \, . 
\ee
 
The dynamics of an expanding Universe  is determined by the Einstein equations,
which relate the expansion  rate to the matter content, specifically to the
energy density $\rho$ and pressure $p$.  For a homogeneous and isotropic
Universe, they reduce to the Friedmann-Robertston-Walker (FRW) equations
\be
\left( {\dot a \over a} \right)^2 - {k\over a^2} = {8 \pi G\over 3 } \rho
\ee
\be
{\ddot a\over a} = - {4 \pi G\over 3} \, (\rho + 3 p) \, .
\ee
These equations can be combined to yield the continuity equation (with Hubble
constant $H = \dot a/a$)
\be \label{cont}
\dot \rho = - 3 H (\rho + p) \, . 
\ee

The third key assumption of standard cosmology is that matter is described by
an ideal gas with an equation of state
\be
p = w \rho \, . 
\ee
For cold matter, pressure is negligible and hence $w = 0$.  From (\ref{cont}) it
follows that
\be
\rho_m (t) \sim a^{-3} (t) \, , 
\ee
where $\rho_m$ is the energy density in cold matter.  For radiation we have $w
= {1/3}$ and hence it follows from (\ref{cont}) that
\be
\rho_r (t) \sim a^{-4} (t) \, , 
\ee
$\rho_r (t)$ being the energy density in radiation.

The three classic observational pillars of standard cosmology are Hubble's law, the existence and black body nature of the nearly isotropic CMB, and the abundances of light elements (nucleosynthesis). These successes are discussed in detail in many textbooks on cosmology, and will therefore not be reviewed here.

It is, however, important to recall two important aspects concerning the thermal history of the early Universe. Since the energy density in radiation redshifts faster than the matter energy density, it follows by working backwards in time from the present data that although the energy density of the Universe is now mostly in cold matter, it was initially dominated by radiation. The transition occurred at a time denoted by $t_{eq}$, the ``time of equal matter and radiation". As will be discussed in Section 3, $t_{eq}$ is the time when perturbations can start to grow by gravitational clustering. The second important time is $t_{rec}$, the ``time of recombination" when photons fell out of equilibrium (since ions and electrons had by then combined to form electrically neutral atoms). The photons of the CMB have travelled without scattering from $t_{rec}$. Their spatial distribution is predicted to be a black body since the cosmological redshift preserves the black body nature of the initial spectrum (simply redshifting the temperature) which was in turn determined by thermal equilibrium. CMB anisotropies probe the density fluctuations at $t_{rec}$. Note that for the usual values of the cosmological parameters, $t_{eq} < t_{rec}$. 

Standard Big Bang cosmology is faced with several important problems.  Only one
of these,  the age problem, is a potential conflict with observations.  The
others which I will focus on here -- the homogeneity, flatness and formation of
structure problems (see e.g. \cite{Guth}) -- are questions which have no answer
within the theory and are therefore the main motivation for the new
cosmological models which will be discussed in later sections of these lecture notes.

The horizon problem is illustrated in Fig. 1.  As is sketched, the comoving
region $\ell_p (t_{rec})$ over which the CMB is observed to be homogeneous  to
better  than one part in $10^4$ is much larger than the comoving forward light
cone $\ell_f (t_{rec})$ at $t_{rec}$, which is the maximal distance over which
microphysical forces could have caused the homogeneity:
\be
\ell_p (t_{rec}) = \int\limits^{t_0}_{t_{rec}} dt \, a^{-1} (t) \simeq 3 \, t_0
\left(1 - \left({t_{rec}\over t_0} \right)^{1/3} \right) 
\ee
\be
\ell_f (t_{rec}) = \int\limits^{t_{rec}}_0 dt \, a^{-1} (t) \simeq 3 \, t^{2/3}_0
\, t^{1/3}_{rec} \, . 
\ee
{}From the above equations it is obvious that $\ell_p (t_{rec}) \gg \ell_f
(t_{rec})$.  Hence, standard cosmology cannot explain the observed isotropy of
the CMB.

\begin{figure}
\begin{center}
\leavevmode
\epsfxsize=6.5cm \epsfbox{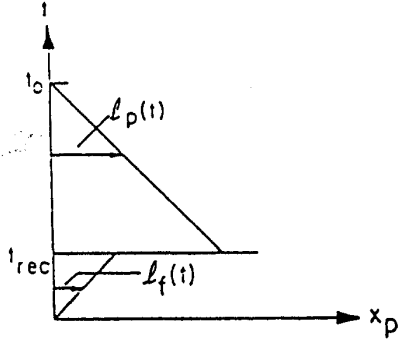}
\caption{
A space-time diagram (physical distance $x_p$ versus time $t$)
illustrating the homogeneity problem: the past light cone $\ell_p (t)$ at the
time $t_{rec}$ of last scattering  is much larger than the forward light cone
$\ell_f (t)$ at $t_{rec}$.}
\end{center}
\end{figure}

In standard cosmology and in an expanding Universe, $\Omega = 1$ is an unstable
fixed point.  This can be seen as follows.  For a spatially flat Universe
$(\Omega = 1)$
\be \label{omega1}
H^2 = {8 \pi G\over 3} \, \rho_c \, , 
\ee
whereas for a nonflat Universe
\be \label{omega2}
H^2 + \varepsilon \, T^2 = {8 \pi G\over 3}  \, \rho \, , 
\ee
with
\be
\varepsilon = {k\over{(aT)^2}} \, . 
\ee
The quantity $\varepsilon$ is proportional to $s^{-2/3}$, where $s$ is the
entropy density.  Hence, in standard cosmology, $\varepsilon$ is constant.
Combining (\ref{omega1}) and (\ref{omega2}) gives
\be \label{omega3}
{\rho - \rho_c\over \rho_c} = {3\over{8 \pi G}} \, {\varepsilon T^2\over
\rho_c} \sim T^{-2} \, . 
\ee
Thus, as the temperature decreases, $\Omega - 1$ increases.  In fact, in order
to explain the present small value of $\Omega - 1 \sim {\cal O} (1)$, the
initial energy density had to be extremely close to critical density.  For
example, at $T = 10^{15}$ GeV, (\ref{omega3}) implies
\be
{\rho - \rho_c\over \rho_c} \sim 10^{-50} \, . 
\ee
What is the origin of these fine tuned initial conditions?  This is the
flatness problem of standard cosmology.

The third of the classic problems of standard cosmological model is the
``formation of structure problem."  Observations indicate that galaxies and
even clusters of galaxies have nonrandom correlations on scales larger than 50
Mpc (see e.g. \cite{CFA,LCRS}).  This scale is comparable to the comoving horizon at
$t_{eq}$.  Thus, if the initial density perturbations were produced much before
$t_{eq}$, the correlations cannot be explained by a causal mechanism.  Gravity
alone is, in general, too weak to build up correlations on the scale of
clusters after $t_{eq}$ (see, however, the explosion scenario of \cite{explosion}).
Hence, the two questions of what generates the primordial density perturbations
and what causes the observed correlations, do not have an answer in the context
of standard cosmology.  This problem is illustrated by Fig. 2.

\begin{figure}
\begin{center}
\leavevmode
\epsfxsize=7.5cm \epsfbox{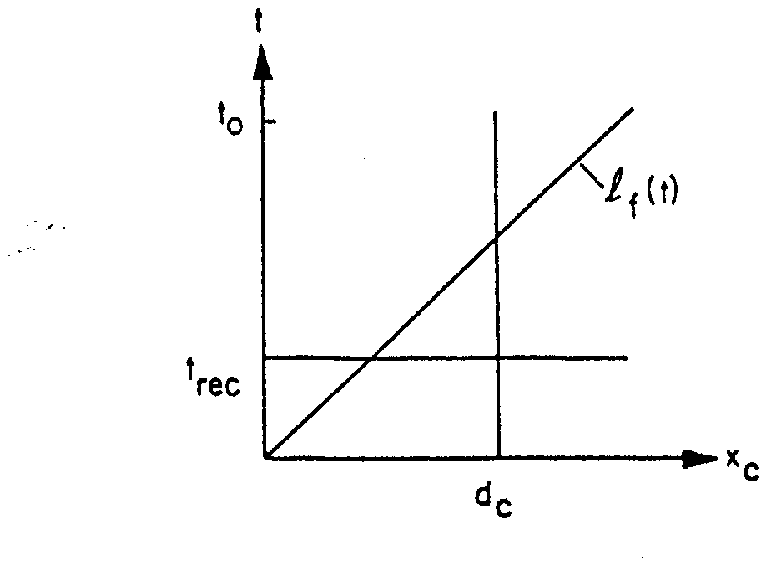}
\caption{A sketch (conformal separation vs. time) of the formation of
structure problem: the comoving separation $d_c$ between two clusters is larger
than the forward light cone at time $t_{eq}$.}
\end{center}
\end{figure}

There are other serious problems of standard cosmology, e.g. the age and the cosmological constant problems. However, to date modern cosmology does not shed any light on these problems, and I will therefore not address them here.

\subsection{Inflationary Universe Scenario}

The idea of inflation$^{\cite{Guth}}$ is very simple (for some early reviews of inflation see e.g. \cite{Linde,GuthBlau,Olive,RB85}).  We assume there is a time
interval beginning at $t_i$ and ending at $t_R$ (the ``reheating time") during
which the Universe is exponentially expanding, i.e.,
\be
a (t) \sim e^{Ht}, \>\>\>\>\> t \epsilon \, [ t_i , \, t_R] 
\ee
with constant Hubble expansion parameter $H$.  Such a period is called  ``de
Sitter" or ``inflationary."  The success of Big Bang nucleosynthesis sets an
upper limit to the time of reheating:
\be
t_R \ll t_{NS} \, , 
\ee
$t_{NS}$ being the time of nucleosynthesis.

\begin{figure}
\begin{center}
\leavevmode
\epsfxsize=10cm \epsfbox{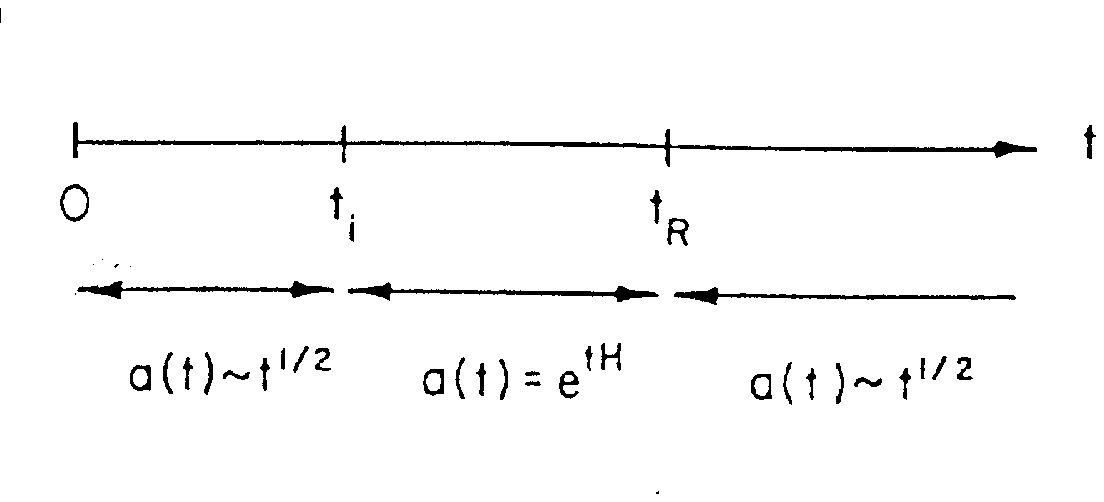}
\caption{The phases of an inflationary Universe. The times
$t_i$ and $t_R$ denote the beginning and end of inflation, respectively.
In some models of inflation, there is no initial radiation dominated FRW
period. Rather, the classical space-time emerges directly in an inflationary
state from some initial quantum gravity state.}
\end{center}
\end{figure}  

The phases of an inflationary Universe are sketched in Fig. 3.  Before the
onset of inflation there are no constraints on the state of the Universe.  In
some models a classical space-time emerges immediately in an inflationary
state, in others there is an initial radiation dominated FRW period.  Our
sketch applies to the second case.  After $t_R$, the Universe is very hot and
dense, and the subsequent evolution is as in standard cosmology.  During the
inflationary phase, the number density of any particles initially in thermal
equilibrium at $t = t_i$ decays exponentially.  Hence, the matter temperature
$T_m (t)$ also decays exponentially.  At $t = t_R$, all of the energy which is
responsible for inflation (see later) is released as thermal energy.  This is a
nonadiabatic process during which the entropy increases by a large factor.   

Fig. 4 is a sketch of how a period of inflation can solve the homogeneity
problem.  $\Delta t = t_R - t_i$  is the period of inflation.  During
inflation, the forward light cone increases exponentially compared to a model
without inflation, whereas the past light cone is not affected for $t \geq
t_R$.  Hence, provided $\Delta t$ is sufficiently large, $\ell_f (t_R)$ will be
greater than $\ell_p (t_R)$. 

\begin{figure}
\begin{center}
\leavevmode
\epsfxsize=12cm \epsfbox{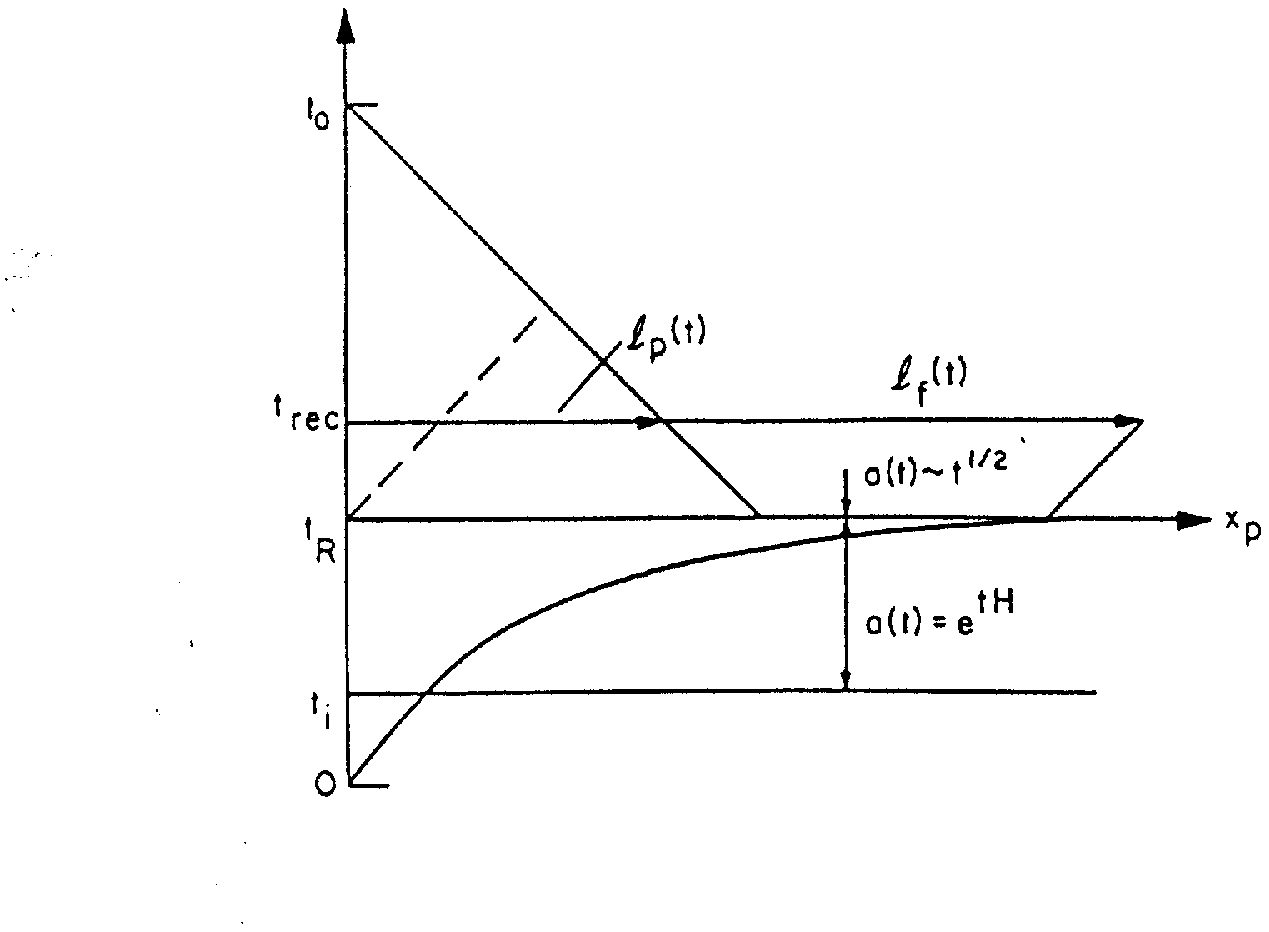}
\caption{
Sketch (physical coordinates vs. time) of the
solution of the homogeneity problem. During inflation, the forward light cone
$l_f(t)$ is expanded exponentially when measured in physical coordinates.
Hence, it does not require many e-foldings of inflation in order that $l_f(t)$
becomes larger than the past light cone at the time of last scattering. The
dashed line is the forward light cone without inflation.}
\end{center}
\end{figure}  

Inflation also can solve the flatness problem$^{\cite{Kazanas,Guth}}$  The key point is
that the entropy density $s$ is no longer constant.  As will be explained
later, the temperatures at $t_i$ and $t_R$ are essentially equal.  Hence, the
entropy increases during inflation by a factor $\exp (3 H \Delta t)$.  Thus,
$\epsilon$ decreases by a factor of $\exp (-2 H \Delta t)$.  Hence, $(\rho - \rho_c) / \rho$ can be of order 1 both at $t_i$ and at the
present time.  In fact, if inflation occurs at all, then rather generically, the theory predicts
that at the present time $\Omega = 1$ to a high accuracy (now $\Omega < 1$
requires  special initial conditions or rather special models$^{\cite{open}}$).

Most importantly, inflation provides a mechanism which in a causal way
generates the primordial perturbations required for galaxies, clusters and even
larger objects.  In inflationary Universe models, the Hubble radius
(``apparent" horizon), $3t$, and the ``actual" horizon (the forward light cone)
do not coincide at late times.  Provided that the duration of inflation is sufficiently long, then (as sketched
in Fig. 5) all scales within our apparent horizon were inside the actual
horizon since $t_i$.  Thus, it is in principle possible to have a casual
generation mechanism for perturbations$^{\cite{Press,Mukh80,Lukash,Sato}}$.

The generation of perturbations is supposed to be due to a causal microphysical
process.  Such processes can only act coherently on length scales smaller than
the Hubble radius $\ell_H (t)$ where
\be
\ell_H (t) = H^{-1} (t) \, . 
\ee
A heuristic way to understand the meaning of $\ell_H (t)$ is to realize that it
is the distance which light (and hence the maximal distance any causal effects)
can propagate in one expansion time.

\begin{figure}
\begin{center}
\leavevmode
\epsfxsize=10cm \epsfbox{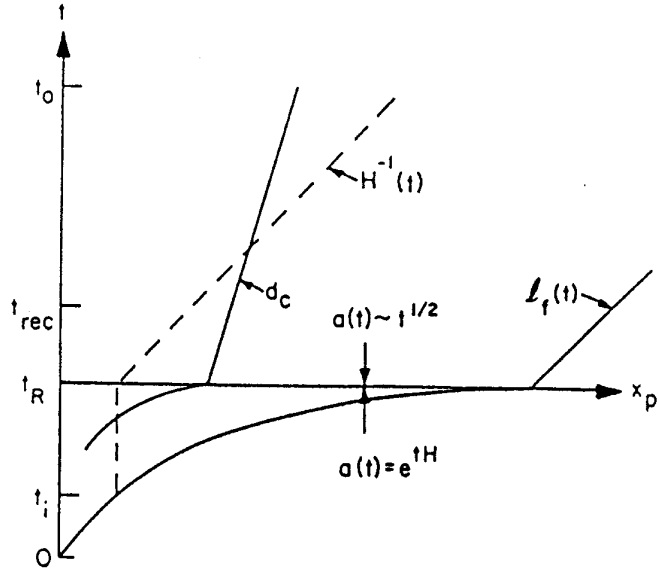}
\caption{
A sketch (physical coordinates vs. time) of the
solution of the formation of structure problem. Provided that the period of
inflation is sufficiently long, the separation $d_c$ between two galaxy
clusters is at all times smaller than the forward light cone. The dashed line
indicates the Hubble radius. Note that $d_c$ starts out smaller than the Hubble
radius, crosses it during the de Sitter period, and then reenters it at late
times.}
\end{center}
\end{figure}

As will be discussed in Chapter 4, the density perturbations produced during
inflation are due to quantum fluctuations in the matter and gravitational
fields$^{\cite{Mukh80,Lukash}}$.  The amplitude of these inhomogeneities corresponds to a
temperature $T_H$
\be
T_H \sim H \, , 
\ee
the Hawking temperature of the de Sitter phase. This implies that at all times
$t$ during inflation, perturbations with a fixed physical wavelength $\sim
H^{-1}$ will be produced. Subsequently, the length of the waves is stretched
with the expansion of space, and soon becomes larger than the Hubble radius.
The phases of the inhomogeneities are random.  Thus, the inflationary Universe
scenario predicts perturbations on all scales ranging from the comoving Hubble
radius at the beginning of inflation to the corresponding quantity at the time
of reheating.  In particular, provided that inflation lasts sufficiently long, perturbations on scales of galaxies and beyond will be generated. Note, however, that it is very dangerous to interpret de Sitter Hawking radiation as thermal radiation. In fact, the equation of state of this ``radiation" is not thermal$^{\cite{RB83}}$.

Obviously, the key question is how to obtain inflation. From the FRW equations, it follows that in order to get exponential increase of the scale factor, the equation of state of matter must be
\be \label{infleos}
p = - \rho  
\ee
This is where the connection with particle physics comes in. The energy density and pressure of a scalar quantum field $\varphi$ are given by
\beq
\rho (\varphi) & = & {1\over 2} \, \dot \varphi^2 + {1\over 2} \,
(\nabla \varphi)^2 + V (\varphi) \label{eos1} \\
p (\varphi) & = & {1\over 2} \dot \varphi^2 - {1\over 6} (\nabla \varphi)^2 - V
(\varphi) \, . \label{eos2}
\eeq
Thus, provided that at some initial time $t_i$
\be \label{incond}
\dot \varphi (\underline{x}, \, t_i) = \nabla \varphi (\underline{x}_i \, t_i)
= 0 
\ee
and
\be
V (\varphi (\underline{x}_i, \, t_i) )  > 0 \, , 
\ee
the equation of state of matter will be (\ref{infleos}).
 
The next question is how to realize the required initial conditions (\ref{incond}) and
to maintain the key constraints
\be
\dot \varphi^2 \ll V (\varphi) \> , \> (\nabla \varphi)^2 \ll V (\varphi)
\ee
for sufficiently long. Various ways of realizing these conditions were put forward, and they gave rise to different models of inflation. I will focus on ``old inflation," ``new inflation"" and ``chaotic
inflation."  There are many other attempts at producing an inflationary
scenario, but there is as of now no convincing realization.

\medskip
\centerline{\bf Old Inflation}
\medskip

The old inflationary Universe model$^{\cite{Guth,GuthTye}}$ is based on a scalar field
theory which undergoes a first order phase transition.  As a toy
model, consider a scalar field theory with the potential $V (\varphi)$
of Figure 6.  This potential has a metastable ``false" vacuum at $\varphi = 0$, whereas the lowest energy state (the ``true" vacuum) is $\varphi = a$. Finite temperature effects$^{\cite{finiteT}}$ lead to extra terms in the finite temperature effective potential which are proportional to $\varphi^2 T^2$ (the resulting finite temperature effective potential is also depicted in Figure 6). Thus, at high temperatures, the energetically preferred state is the false vacuum state. Note that this is only true if $\varphi$ is in thermal equilibrium with the other fields in the system.

\begin{figure}
\begin{center}
\leavevmode
\epsfxsize=10cm \epsfbox{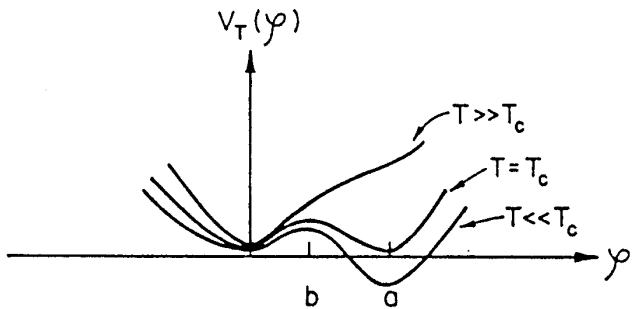}
\caption{
The finite temperature effective potential
in a theory with a first order phase transition.}
\end{center}
\end{figure}
     
\par
For fairly general initial conditions, $\varphi (x)$ is trapped in the
metastable state $\varphi = 0$ as the Universe cools below the
critical temperature $T_c$.  As the Universe expands further, all
contributions to the energy-momentum tensor $T_{\mu \nu}$ except for
the contribution
\be
T_{\mu \nu} \sim V(\varphi) g_{\mu \nu} 
\ee
redshift.  Hence, provided that the potential $V(\varphi)$ is shifted upwards such that $V(a) = 0$, then the equation of state in the false vacuum approaches $p = - \rho$, and
inflation sets in. After a period $\Gamma^{-1}$, where $\Gamma$ is the tunnelling rate, bubbles of $\varphi = a$ begin to nucleate$^{\cite{decay}}$ in a sea of false
vacuum $\varphi = 0$. Inflation lasts until the false vacuum decays.
During inflation, the Hubble constant is given by
\be
H^2 = {8 \pi G\over 3} \, V (0) \, . 
\ee
Note that the condition $V(a) = 0$, which looks rather unnatural, is required to
avoid a large cosmological constant today (none of the present inflationary Universe
models manages to circumvent or solve the cosmological constant problem).
 
It was immediately realized that old inflation has a serious ``graceful exit"
problem$^{\cite{Guth,GuthWein}}$.  The bubbles nucleate after inflation with radius $r \ll
2t_R$ and would today be much smaller than our apparent horizon.  Thus, unless
bubbles percolate, the model predicts extremely large inhomogeneities inside
the Hubble radius, in contradiction with the observed isotropy of the
microwave background radiation.
\par
For bubbles to percolate, a sufficiently large number must be produced so that
they collide and homogenize over a scale larger than the present Hubble
radius.  However, with exponential expansion, the volume between bubbles
expands
exponentially whereas the volume inside bubbles expands only with a low power.
This prevents percolation.

\medskip
\centerline{\bf New Inflation}
\medskip

Because of the graceful exit problem, old inflation never was considered to be
a viable cosmological model.  However, soon after the seminal paper by
Guth, Linde$^{\cite{Linde82}}$ and independently Albrecht and Steinhardt$^{\cite{AS82}}$ put
forwards a modified scenario, the New Inflationary Universe.

The starting point is a scalar field theory with a double well potential which
undergoes a second order phase transition (Fig. 7).  $V(\varphi)$ is
symmetric and $\varphi = 0$ is a local maximum of the zero temperature
potential.  Once again, it was argued that finite temperature effects confine
$\varphi(x)$ to values near $\varphi = 0$ at temperatures $T \geq
T_c$.  For $T < T_c$, thermal fluctuations trigger the instability of $\varphi
(x) = 0$ and $\varphi (x)$ evolves towards either of the global minima at $\varphi = \pm \sigma$ by the classical equation of motion
\be \label{eom}
\ddot \varphi + 3 H \dot \varphi - a^{-2} \bigtriangledown^2 \varphi = -
V^\prime (\varphi)\, .  
\ee

Within a fluctuation region, 
$\varphi(x)$ will be homogeneous. In such a region, we can  neglect the spatial gradient terms in Eq. (\ref{eom}).  Then, from (\ref{eos1}) and (\ref{eos2}) we can read off the
induced equation of state.  The condition for inflation is
\be
\dot \varphi^2 \ll V (\varphi)\, , 
\ee
i.e.~ slow rolling.
Often, the  ``slow rolling" approximation is made to find solutions of
(\ref{eom}).
This consists of dropping the $\ddot \varphi$ term.   

\begin{figure}
\begin{center}
\leavevmode
\epsfxsize=8cm \epsfbox{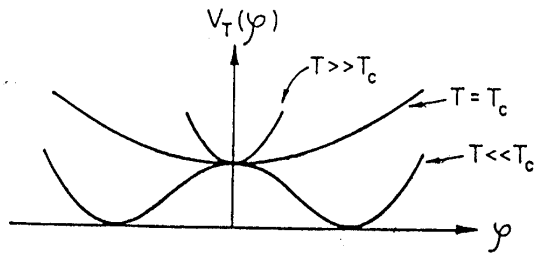}
\caption{ 
The finite temperature effective potential
in a theory with a second order phase transition.}
\end{center}
\end{figure}
 
There is no graceful exit problem in the new inflationary Universe.  Since the
fluctuation domains are established before the onset of inflation,
any boundary walls will be inflated outside the present Hubble radius.
\par
Let us, for the moment, return to the general features of the new inflationary
Universe scenario.  At the time $t_c$ of the phase transition, $\varphi (t)$
will start to move from near $\varphi = 0$ towards either $\pm \sigma$ as
described by the classical equation of motion, i.e.~ (\ref{eom}).  At or soon after
$t_c$, the energy-momentum tensor of the Universe will start to be dominated
by $V(\varphi)$, and inflation will commence.  $t_i$ shall denote the time of
the onset of inflation.  Eventually, $\phi (t)$ will reach large values for which
nonlinear effects become important.  The time at which this occurs is $t_B$.
For $t > t_B \, , \, \varphi (t)$ rapidly accelerates, reaches $\pm \sigma$,
overshoots and starts oscillating about the global minimum of $V (\varphi)$.
The amplitude of this oscillation is damped by the expansion of the Universe
and (predominantly) by the coupling of $\varphi$ to other fields.  At time
$t_R$,
the energy in $\varphi$ drops below the energy of the thermal bath of
particles produced during the period of oscillation.
\par
The evolution of $\varphi (t)$ is sketched in Fig. 8.  The time period
between $t_B$ and $t_R$ is called the reheating period and is usually short
compared to the Hubble expansion time. For $t > t_R$, the Universe is again radiation dominated.

\begin{figure}
\begin{center}
\leavevmode
\epsfxsize=13cm \epsfbox{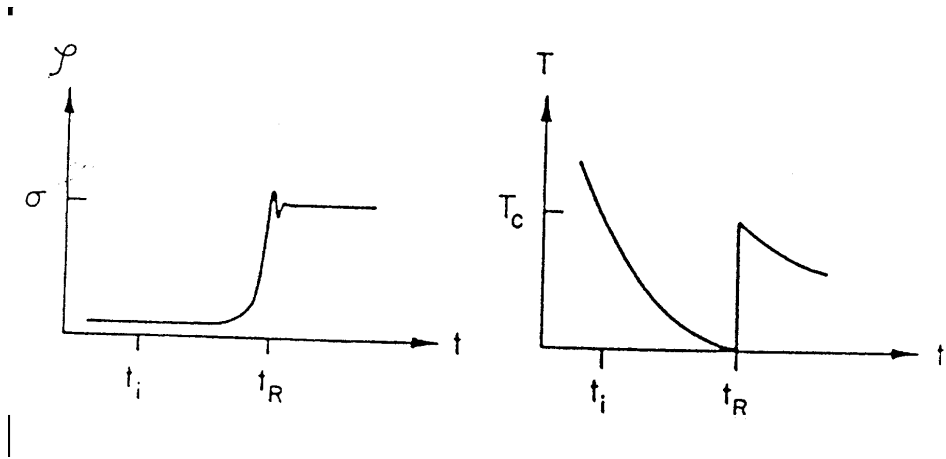}
\caption{ 
Evolution of $\varphi (t)$ and $T (t)$ in the new
inflationary Universe.}
\end{center}
\end{figure} 
 
In order to obtain inflation, the potential $V(\varphi)$ must be very flat near the false vacuum at $\varphi = 0$. This can only be the case if all of the coupling constants appearing in the potential are small. However, this implies that the  $\varphi$ cannot be in thermal equilibrium at early times, which would be required to localize $\varphi$ in the false vacuum. In the absence of thermal equilibrium, the initial conditions for $\varphi$ are only constrained by requiring that the total energy density in $\varphi$ not exceed the total energy density of the Universe. Most of the phase space of these initial conditions lies at values of $| \varphi | >> \sigma$. This leads to the ``chaotic" inflation scenario$^{\cite{Linde83}}$.

\medskip  
\centerline{\bf Chaotic Inflation}
\medskip

Consider a region in space where at the initial time $\varphi (x)$
is very large, homogeneous and static.  In this case, the energy-momentum tensor will be
immediately dominated by the large potential energy term and induce an
equation of state $p \simeq - \rho$ which leads to inflation.  Due to the
large Hubble damping term in the scalar field equation of motion, $\varphi
(x)$ will only roll very slowly towards $\varphi = 0$.  The
kinetic energy contribution to $T_{\mu \nu}$ will remain small, the spatial
gradient contribution will be exponentially suppressed due to the expansion of
the Universe, and thus inflation persists. Note that in contrast to old and new inflation,
no initial thermal bath is required.  Note also that the precise form of
$V(\varphi)$ is irrelevant to the mechanism.  In particular, $V(\varphi)$ need
not be a double well potential.  This is a significant advantage, since for
scalar fields other than Higgs fields used for spontaneous symmetry breaking,
there is no particle physics motivation for assuming a double well potential,
and since the inflaton (the field which gives rise to inflation) cannot be a
conventional Higgs field due to the severe fine tuning constraints.
\par
The field and temperature evolution in a chaotic inflation model is similar to what is depicted in Figure 8, except that $\varphi$ is rolling towards the true vacuum at $\varphi = \sigma$ from the direction of large field values.

Chaotic inflation is a much more radical departure from standard cosmology than old and new inflation. In the latter, the inflationary phase can be viewed as a short phase of exponential expansion bounded at both ends by phases of radiation domination. In chaotic inflation, a piece of the Universe emerges with an inflationary equation of state immediately after the quantum gravity (or string) epoch.

The chaotic inflationary Universe scenario has been developed in great detail (see e.g. \cite{Linde94} for a recent review). One important addition is the inclusion of stochastic noise$^{\cite{Starob87}}$ in the equation of motion for $\varphi$ in order to take into account the effects of quantum fluctuations. It can in fact be shown that for sufficiently large values of $|\varphi|$, the stochastic force terms are more important than the classical relaxation force $V^\prime(\varphi)$. There is equal probability for the quantum fluctuations to lead to an increase or decrease of $|\varphi|$. Hence, in a substantial fraction of comoving volume, the field $\varphi$ will climb up the potential. This leads to the conclusion that chaotic inflation is eternal. At all times, a large fraction of the physical space will be inflating. Another consequence of including stochastic terms is that on large scales (much larger than the present Hubble radius), the Universe will look extremely inhomogeneous.

\subsection{Problems of Inflationary Cosmology}

In spite of its great success at resolving some of the problems of standard cosmology and of providing a causal, predictive theory of structure formation, there are several important unresolved conceptual problems in inflationary cosmology. I will focus on three of these problems, the cosmological constant mystery, the fluctuation problem, and the dynamics of reheating.

\medskip
\centerline{\bf Cosmological Constant Problem}
\medskip

Since the cosmological constant acts as an effective energy density, its value is bounded from above by the present energy density of the Universe. In Planck units, the constraint on the effective cosmological constant $\Lambda_{eff}$ is
(see e.g. \cite{cosmorev})
\be
{{\Lambda_{eff}} \over {m_{pl}^4}} \, \le \, 10^{- 122} \, .
\ee
This constraint applies both to the bare cosmological constant and to any matter contribution which acts as an effective cosmological constant.

The true vacuum value of the potential $V(\varphi)$ acts as an effective cosmological constant. Its value is not constrained by any particle physics requirements (in the absence of special symmetries). The cosmological constant problem is thus even more accute in inflationary cosmology than it usually is. The same unknown mechanism which must act to shift the potential (see Figure 6) such that inflation occurs in the false vacuum must also adjust the potential to vanish in the true vacuum. 

Supersymmetric theories may provide a resolution of this problem, since unbroken supersymmetry forces $V(\varphi) = 0$ in the supersymmetric vacuum. However, supersymmetry breaking will induce a nonvanishing $V(\varphi)$ in the true vacuum after supersymmetry breaking.

We may therefore be forced to look for realizations of inflation which do not make use of scalar fields. There are several possibilities. It is possible to obtain inflation in higher derivative gravity theories. In fact, the first
model with exponential expansion of the Universe was obtained$^{\cite{Starob}}$ in an $R^2$ gravity theory. The extra degrees of freedom associated with the higher derivative terms act as scalar fields with a potential which automatically vanishes in the true vacuum. For some recent work on higher derivative gravity inflation see also \cite{MB92}. 

Another way to obtain inflation is by making use of condensates (see \cite{Ball} and \cite{Parker} for different approaches to this problem). An additional motivation for following this route to inflation is that the symmetry breaking mechanisms observed in nature (in condensed matter systems) are induced by the formation of condensates such as Cooper pairs. Again, in a model of condensates there is no freedom to add a constant to the effective potential.

The main problem of studying the possibility of obtaining inflation using condensates is that the quantum effects which determine the theory are highly nonperturbative. In particular, the effective potential written in terms of a condensate $\cond$ does not correspond to a renormalizable theory and will in general$^{\cite{ARZ}}$ contain terms of arbitrary power in $\cond$. However (see \cite{BZ96}), one may make progress by assuming certain general properties of the effective potential.

Let us$^{\cite{BZ96}}$ consider a theory in which at some time $t_i$ a condensate $\cond$ forms, i.e. $\cond = 0$ for $t < t_i$ and $\cond \neq 0$ for $t > t_i$. The expectation value of the Hamiltonian $H$ written in terms of the condensate $\cond$ contains terms of arbitrary powers of $\cond$:
\be
\langle H \rangle \, = \, \sum_{n} (-1)^n {n!} a_n \cond^n \, .
\ee
We summarize our ignorance of the nonperturbative physics in the assumption that the resulting series is asymptotic, and in particular Borel summable, with coefficients $a_n \propto 1$. In this case, we can resum the series to obtain$^{\cite{BZ96}}$
\be \label{effpot}
\langle H \rangle \, = \, \int_0^{\infty} {{f(t) dt} \over {t (t m_{pl} + \cond)}} e^{- 1/t} \, ,
\ee
where the function $f(t)$ is related to the coefficients $a_n$ via
\be
a_n \, = \, {1 \over {n!}} \int_0^{\infty} dt f(t) t^{-n - 2} e^{- 1/t} \, .
\ee

The expectation value of the Hamiltonian $\langle H \rangle$ can be interpreted as the effective potential $V_{eff}$ of this theory. The question is under which conditions this potential gives rise to inflation. If we regard $\cond$ as a classical field (i.e. neglect the ultraviolet and infrared divergences of the theory), then the dynamics of the model can be read off directly from (\ref{effpot}), with initial conditions for $\cond$ at the time $t_i$ close to $\cond = 0$. It is easy to check that rather generically, the conditions required to have slow rolling of $\varphi$, namely
\be
V^\prime m_{pl} \, << \, \sqrt{48 \pi} V 
\ee
\be
V^{\prime \prime} m_{pl}^2 \, << \, 24 \pi V \, ,
\ee
are satisfied. However, since the potential decays only slowly at large values of $\cond$ and since there is no true vacuum state at finite values of $\cond$, the slow rolling conditions are satisfied for all times. In this case, inflation would never end - an obvious cosmological disaster.

However, $\cond$ is not a classical scalar field but the expectation value of a condensate operator. Thus, we have to worry about diverging contributions to this expectation value. In particular, in a theory with symmetry breaking there will often be massless excitations which will give rise to infrared divergences. It is necessary to introduce an infrared cutoff energy $\varepsilon$ whose value is determined in the context of cosmology by the Hubble expansion rate. Note in particular that this cutoff is time-dependent. Effectively, we thus have a theory of two scalar fields $\cond$ and $\varepsilon$. 
In this case, the first of the slow rolling conditions becomes (if $\varepsilon$ is expressed in Planck units)
\be
\dot{\varepsilon}^2 m_{pl}^2 + \dot{\varphi}^2 \, << \, 2 V \, .
\ee

The infrared cutoff changes the form of the effective potential. We assume that this change can be modelled by replacing $\cond$ by $\cond / \varepsilon$. If we
(following \cite{Woodard}) take the infrared cutoff to be
\be \label{ansatz}
\varepsilon(t) = {{H(0)} \over {m_{pl}}} [1 - a (Ht)^p] \, ,
\ee
where $0 < a << 1$ and $p$ is an integer and the time at the beginning of the rolling has been set to $t = 0$, then it can be shown$^{\cite{BZ96}}$ that an period of inflation with a graceful exit is realized. After the
condensate $\cond$ starts rolling at $\cond \sim 0$, inflation will commence. As inflation proceeds, $\varepsilon(t)$ will slowly grow and will eventually dominate the energy functional, signaling an end of the inflationary period. From (\ref{ansatz}) it follows that inflation lasts until $a^{1/p} H t = 1$. 

This analysis demonstrates that it is in principle possible to obtain inflation from condensates. However, the model must be studied in much more detail before we can determine whether it gives a realization of inflation which is free of problems.

\medskip
\centerline{\bf Fluctuation Problem}
\medskip

A generic problem for all realizations of inflation studied up to now concerns the amplitude of the density perturbations which are induced by quantum fluctuations during the period of exponential expansion. From the amplitude of CMB anisotropies measured by COBE, and from the present amplitude of density inhomogeneities on scales of clusters of galaxies, it follows that the amplitude of the mass fluctuations ${\delta M} / M$ on a length scale given by the comoving wavenumber $k$ at the time $t_H(k)$ when that scale crosses the Hubble radius in the FRW period is 
\be \label{obs}
{{\delta M} \over M} (k, t_H(k)) \, \propto \, 10^{-5} \, .
\ee

The generation and evolution of fluctuations will be discussed in detail in Section 3. The perturbations arise during inflation as quantum excitations. Their amplitude at the time $t_i(k)$ when the scale $k$ leaves the Hubble radius during inflation is given by
\be \label{inpert}
{{\delta M} \over M} (k, t_i(k)) \, \simeq \, {{V^{\prime} \delta \varphi} \over \rho}|_{t_i(k)} \, ,
\ee
where $\delta \varphi$ is given by the amplitude of the quantum fluctuation of $\delta \varphi(k)$ (note that this is a momentum space quantity). While the scale $k$ is outside of the Hubble radius, the fluctuation amplitude grows by general relativistic gravitational effects. The amplitudes at $t_i(k)$ and $t_H(k)$ are related by
\be \label{amplpert}
{{\delta M} \over M} (K, t_H(k)) \, \simeq \, {1 \over {1 + p / \rho}}|_{t_i(k)}
{{\delta M} \over M} (k, t_i(k)) 
\ee
(see e.g. \cite{MFB92}). Combining (\ref{inpert}) and (\ref{amplpert}) and working out the result for the potential
\be
V(\varphi) \, = \, \lambda \varphi^4
\ee
we obtain the result$^{\cite{Mukh81,flucts,BST}}$
\be
{{\delta M} \over M} (K, t_H(k)) \, \simeq 10^2 \lambda^{1/2} \, .
\ee
Thus, in order to agree with the observed value (\ref{obs}), the coupling constant $\lambda$ must be extremely small:
\be \label{fluctconstr}
\lambda \, \leq \, 10^{-12} \, .
\ee

It has been shown in \cite{Freese} that the above conclusion is generic, at least for models in which inflation is driven by a scalar field. In order that inflation does not produce a too large amplitude of the spectrum of perturbations, a dimensionless number appearing in the potential must be set to a very small value. A possible resolution of this problem will be mentioned in the following subsection.

\medskip
\centerline{\bf Reheating Problem}
\medskip

A question which has recently received a lot of attention and will be discussed in greater detail in one of the following subsections is the issue of reheating in inflationary cosmology. The question concerns the energy transfer between the inflaton and matter fields which is supposed to take place at the end of inflation (see Fig. 8). 

According to either new inflation or chaotic inflation, the dynamics of the inflaton leads first to a transfer of energy from potential energy of the inflaton to kinetic energy. After the period of slow rolling, the inflaton $\varphi$ begins to oscillate about the true minimum of $V(\varphi)$. Quantum mechanically, the state of homogeneous oscillation corresponds to a coherent state. Any coupling of $\varphi$ to other fields (and even self coupling terms of $\varphi$) will lead to a decay of this state. This corresponds to the particle production. The produced particles will be relativistic, and thus at the conclusion of the reheating period a radiation dominated Universe will emerge.

The key questions are by what mechanism and how fast the decay of the coherent state takes place. It is important to determine the temperature of the produced particles at the end of the reheating period. The answers are relevant to many important questions regarding the post-inflationary evolution. For example, it is important to know whether the temperature after reheating is high enough to allow GUT baryogenesis and the production of GUT-scale topological defects. In supersymmetric models, the answer determines the predicted abundance of gravitinos and other moduli fields.

Recently, there has been a complete change in our understanding of reheating. This topic will be discussed in detail below.

\subsection{Inflation and Nonsingular Cosmology}

The question we wish to address in this subsection is whether it is 
possible to construct a class of effective actions for gravity which 
have improved singularity properties and which predict inflation, 
with the constraint that they give the correct low curvature limit.
Since Planck scale physics will generate corrections to the Einstein action, it is quite reasonable to consider higher derivative gravity models.

What follows is a summary of recent work$^{\cite{MB92}}$ in which we have 
constructed an effective action for gravity in which all solutions 
with sufficient symmetry are nonsingular.  The theory is a higher 
derivative modification of the Einstein action, and is obtained by 
a constructive procedure well motivated in analogy with the analysis 
of point particle motion in special relativity.  The resulting theory 
is asymptotically free in a sense which will be specified below.  

Our aim is to construct a theory with the property that the metric 
$g_{\mu\nu}$ approaches the de Sitter metric $g_{\mu\nu}^{DS}$, a 
metric with maximal symmetry which admits a geodesically complete and 
nonsingular extension, as the curvature $R$ approaches the Planck 
value $R_{pl}$.  Here, $R$ stands for any curvature invariant.  
Naturally, from our classical considerations, $R_{pl}$ is a free 
parameter.  However, if our theory is connected with Planck scale 
physics, we expect $R_{pl}$ to be set by the Planck scale.

If successful, the above construction will have some very appealing 
consequences.  Consider, for example, a collapsing spatially 
homogeneous Universe.  According to Einstein's theory, this Universe 
will collapse in finite proper time to a final ``big crunch" singularity (top left Penrose diagram of Figure 9). 
In our theory, however, the Universe will approach a de Sitter model as 
the curvature increases.  If the 
Universe is closed, there will be a de Sitter bounce followed by 
re-expansion (bottom left Penrose diagram in Figure 9).  Similarly, in our theory spherically 
symmetric vacuum  solutions would be nonsingular, i.e., black holes 
would have no singularities in their centers.  The structure of a 
large black hole would be unchanged compared to what is predicted by 
Einstein's theory (top right, Figure 9) outside and even slightly inside the horizon, since 
all curvature 
invariants are small in those regions.  However, for $r \rightarrow 0$ 
(where $r$ is the radial Schwarzschild coordinate), the solution 
changes and approaches a de Sitter solution (bottom right, Figure 9).  This 
would have interesting consequences for the black hole information 
loss problem.

\begin{figure}
\begin{center}
\leavevmode
\epsfxsize=6in \epsfbox{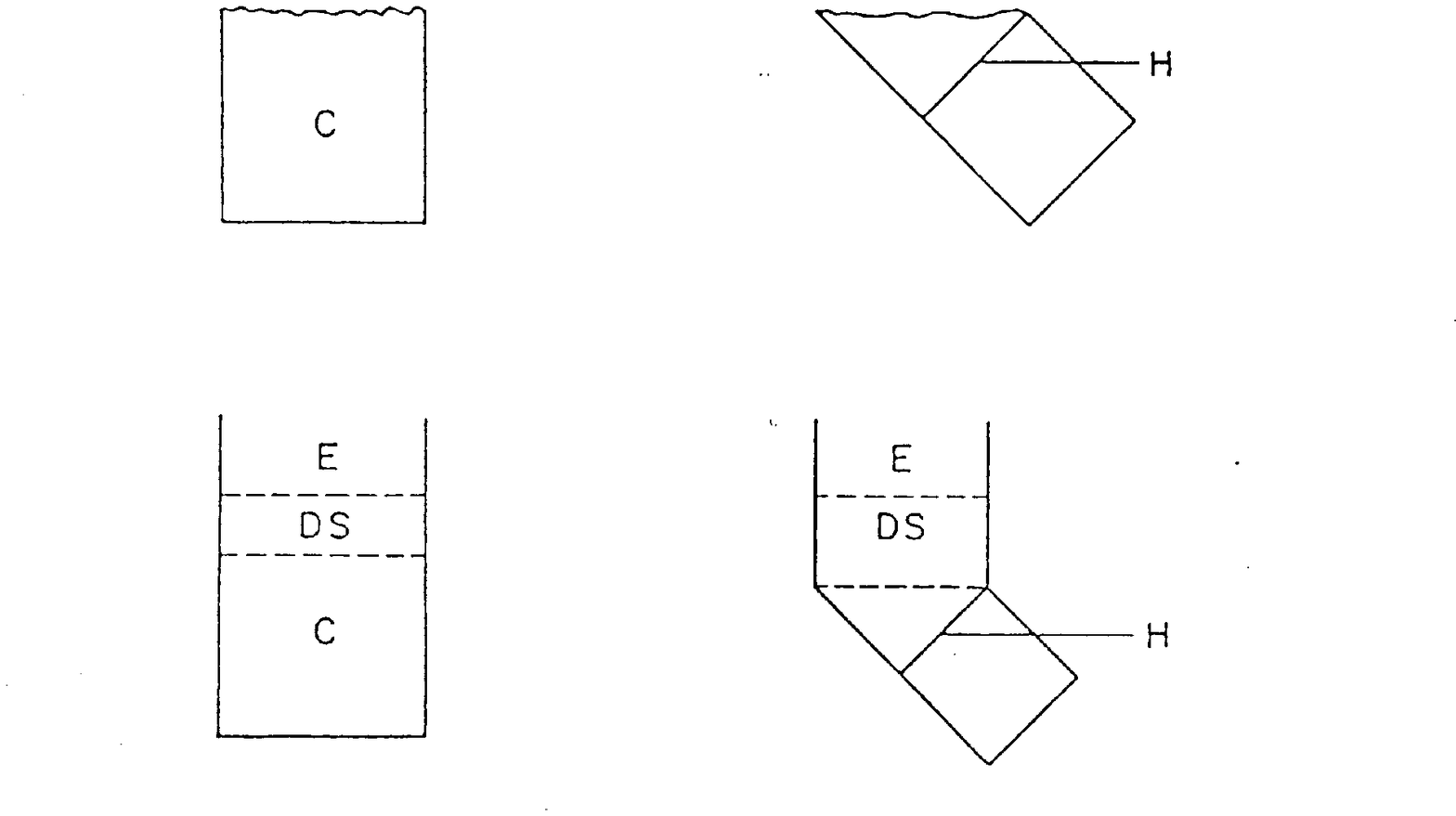}
\caption{
Penrose diagrams for collapsing Universe (left) and black hole (right) in Einstein's theory (top) and in the nonsingular Universe (bottom). C, E, DS and H stand for contracting phase, expanding phase, de Sitter phase and horizon, respectively, and wavy lines indicate singularities.}
\end{center}
\end{figure}

To motivate our effective action construction, we turn to a well known 
analogy, point particle motion in the theory of special relativity.

\medskip
\centerline{\bf An Analogy}
\medskip

The transition from the Newtonian theory of point particle motion to 
the special relativistic theory transforms a theory with no bound on 
the velocity into one in which there is a limiting velocity, the speed 
of light $c$ (in the following we use units in which $\hbar = c = 1$).  
This transition can be obtained$^{\cite{MB92}}$ by starting with the action of a point particle with world line $x(t)$:
\be
S_{\rm old} = \int dt {1\over 2} \dot x^2 \, , 
\ee
introducing$^{\cite{Alt}}$ a Lagrange multiplier field $\varphi$which couples to $\dot x^2$, the quantity to be made finite, and which has a potential 
$V(\varphi)$. The new action is
\be
S_{\rm new} = \int dt \left[ {1\over 2} \dot x^2 + \varphi \dot x^2 - 
V (\varphi) \right] \, . 
\ee
From the constraint equation
\be
\dot x^2 = {\partial V\over{\partial \varphi}} \, , 
\ee
it follows that $\dot x^2$ is limited provided $V(\varphi)$ increases 
no faster than linearly in $\varphi$ for large $|\varphi|$.  The small 
$\varphi$ asymptotics of $V(\varphi)$ is determined by demanding that 
at low velocities the correct Newtonian limit results:
\beq 
V (\varphi) \, \sim \, \varphi^2 \>\,\,\, & {\rm as} & \> |\varphi| 
\rightarrow 0 \, , \label{ascond1} \\
V (\varphi) \sim \varphi \>\,\,\, & {\rm as}  & \> |\varphi| \rightarrow \infty 
\, . \label{ascond2} 
\eeq
Choosing the simple interpolating potential
\be
V (\varphi) \, = \, {2 \varphi^2\over{1 + 2 \varphi}} \, , 
\ee
the Lagrange multiplier can be integrated out, resulting in the well-known
action
\be
S_{\rm new} \, = \, {1\over 2} \int dt \sqrt{1 - \dot x^2} 
\ee
for point particle motion in special relativity.

\medskip
\centerline{\bf Construction}
\medskip

 Our procedure for obtaining a nonsingular Universe theory$^{\cite{MB92}}$ is based 
on generalizing the above Lagrange multiplier construction to gravity.  
Starting from the Einstein action, we can introduce a Lagrange 
multiplier $\varphi_1$ coupled to the Ricci scalar $R$ to obtain a 
theory with limited $R$:
\be
S \, = \, \int d^4 x \sqrt{-g} (R + \varphi_1 \, R + V_1 (\varphi_1) ) \, , 
\ee
where the potential $V_1 (\varphi_1)$ satisfies the asymptotic 
conditions (\ref{ascond1}) and (\ref{ascond2}).

However, this action is insufficient to obtain a nonsingular gravity 
theory.  For example, singular solutions of the Einstein equations 
with $R=0$ are not effected at all.  The minimal requirements for a 
nonsingular theory is that \underbar{all} curvature invariants remain 
bounded and the space-time manifold is geodesically complete.  
Implementing the limiting curvature hypothesis$^{\cite{LCH}}$, these conditions 
can be reduced to more manageable ones.  First, we choose one 
curvature invariant $I_1 (g_{\mu\nu})$ and demand that it be 
explicitely bounded, i.e., $|I_1| < I_1^{pl}$, where $I_1^{pl}$ is the 
Planck scale value of $I_1$.  In a second step, we demand that as $I_1 
(g_{\mu\nu})$ approaches $I_1^{pl}$, the metric $g_{\mu\nu}$ approach 
the de Sitter metric $g^{DS}_{\mu\nu}$, a definite nonsingular metric 
with maximal symmetry.  In this case, all curvature invariants are 
automatically bounded (they approach their de Sitter values), and the 
space-time can be extended to be geodesically complete.

Our approach is to implement the second step of the above procedure by 
another Lagrange multiplier construction$^{\cite{MB92}}$.  We look for a curvature 
invariant $I_2 (g_{\mu\nu})$ with the property that 
\be
I_2 (g_{\mu\nu}) = 0 \>\> \Leftrightarrow \>\> g_{\mu\nu} = 
g^{DS}_{\mu\nu} \, , 
\ee
introduce a second Lagrange multiplier field $\varphi_2$ which couples 
to $I_2$ and choose a potential $V_2 (\varphi_2)$ which forces $I_2$ 
to zero at large $|\varphi_2|$:
\be
S \, = \, \int d^4  x \sqrt{-g} [ R + \varphi_1 I_1 + V_1 (\varphi_1) + 
\varphi_2 I_2 + V_2 (\varphi_2) ] \, , 
\ee
with asymptotic conditions (\ref{ascond1}) and (\ref{ascond2}) for $V_1 (\varphi_1)$ and conditions
\beq
V_2 (\varphi_2) & \sim & {\rm const} \>\>\,\,\, {\rm as} \> | 
\varphi_2 | \rightarrow \infty \\
V_2 (\varphi_2) & \sim & \varphi^2_2 \>\>\,\,\, {\rm as} \> |\varphi_2 | 
\rightarrow 0 \, , 
\eeq
for $V_2 (\varphi_2)$.  The first constraint forces $I_2$ to zero, the 
second is required in order to obtain the correct low curvature limit.

These general conditions are reasonable, but not sufficient in order 
to obtain a nonsingular theory.  It must still be shown that all 
solutions are well behaved, i.e., that they asymptotically reach the 
regions $|\varphi_2| \rightarrow \infty$ of phase space (or that 
they can be controlled in some other way).  This must be done for a 
specific realization of the above general construction.

\medskip
\centerline{\bf Specific Model}
\medskip

At the moment we are only able to find an invariant $I_2$ which 
singles out de Sitter space (by demanding $I_2 = 0$) provided we assume 
that the metric has special symmetries.  The choice
\be
I_2 = (4  R_{\mu\nu} R^{\mu\nu} - R^2 + C^2)^{1/2} \, , 
\ee
singles out the de Sitter metric among all homogeneous and isotropic 
metrics (in which case adding $C^2$, the Weyl tensor square, is 
superfluous), all homogeneous and anisotropic metrics, and all 
radially symmetric metrics.

We choose the action$^{\cite{MB92,BMS93}}$
\be
S \, = \, \int d^4 x \sqrt{-g} \left[ R + \varphi_1 R - (\varphi_2 + 
{3\over{\sqrt{2}}} \varphi_1) I_2^{1/2} + V_1 (\varphi_1) + V_2 
(\varphi_2) \right] 
\ee
with
\be
V_1 (\varphi_1) \, = \, 12 \, H^2_0 {\varphi^2_1\over{1 + \varphi_1}} \left( 1 
- {\ln (1 + \varphi_1)\over{1 + \varphi_1}} \right) 
\ee
\be
V_2 (\varphi_2) \, = \, - 2 \sqrt{3} \, H^2_0 \, {\varphi^2_2\over{1 + 
\varphi^2_2}} \, . 
\ee

The general equations of motion resulting from this action are quite 
messy.  However, when restricted to homogeneous and isotropic metrics 
of the form
\be
ds^2 \, = \, dt^2 - a (t)^2 (dx^2 + dy^2 + dz^2) \, , 
\ee
the equations are fairly simple.  With $H = \dot a / a$, the two 
$\varphi_1$ and $\varphi_2$ constraint equations are
\be \label{nseom1}
H^2 \, = \, {1\over{12}} V^\prime_1 
\ee
\be \label{nseom2}
\dot H \, = \, - {1\over{2\sqrt{3} }} V^\prime_2 \, , 
\ee
and the dynamical $g_{00}$ equation becomes
\be \label{nseom3}
3 (1 - 2 \varphi_1) H^2 + {1\over 2} (V_1 + V_2) \, = \, \sqrt{3} H (\dot 
\varphi_2 + 3 H \varphi_2) \, . 
\ee
The phase space of all vacuum configurations is the half plane $\{ 
(\varphi_1 \geq 0, \, \varphi_2) \}$.  Equations (\ref{nseom1}) and (\ref{nseom2}) 
can be used to express $H$ and $\dot H$ in terms of $\varphi_1$ and 
$\varphi_2$.  The remaining dynamical equation (\ref{nseom3}) can then be recast as
\be
{d \varphi_2\over{d \varphi_1}} \,  = \, - {V_1^{\prime\prime}\over{4 
V^\prime_2}} \, \left[ - \sqrt{3} \varphi_2 + (1 - 2\varphi_1) - 
{2\over{V^\prime_1}} (V_1 + V_2) \right] \, . 
\ee
The solutions can be studied analytically in the asymptotic regions 
and numerically throughout the entire phase space.

\begin{figure}
\begin{center}
\leavevmode
\epsfxsize=6in \epsfbox{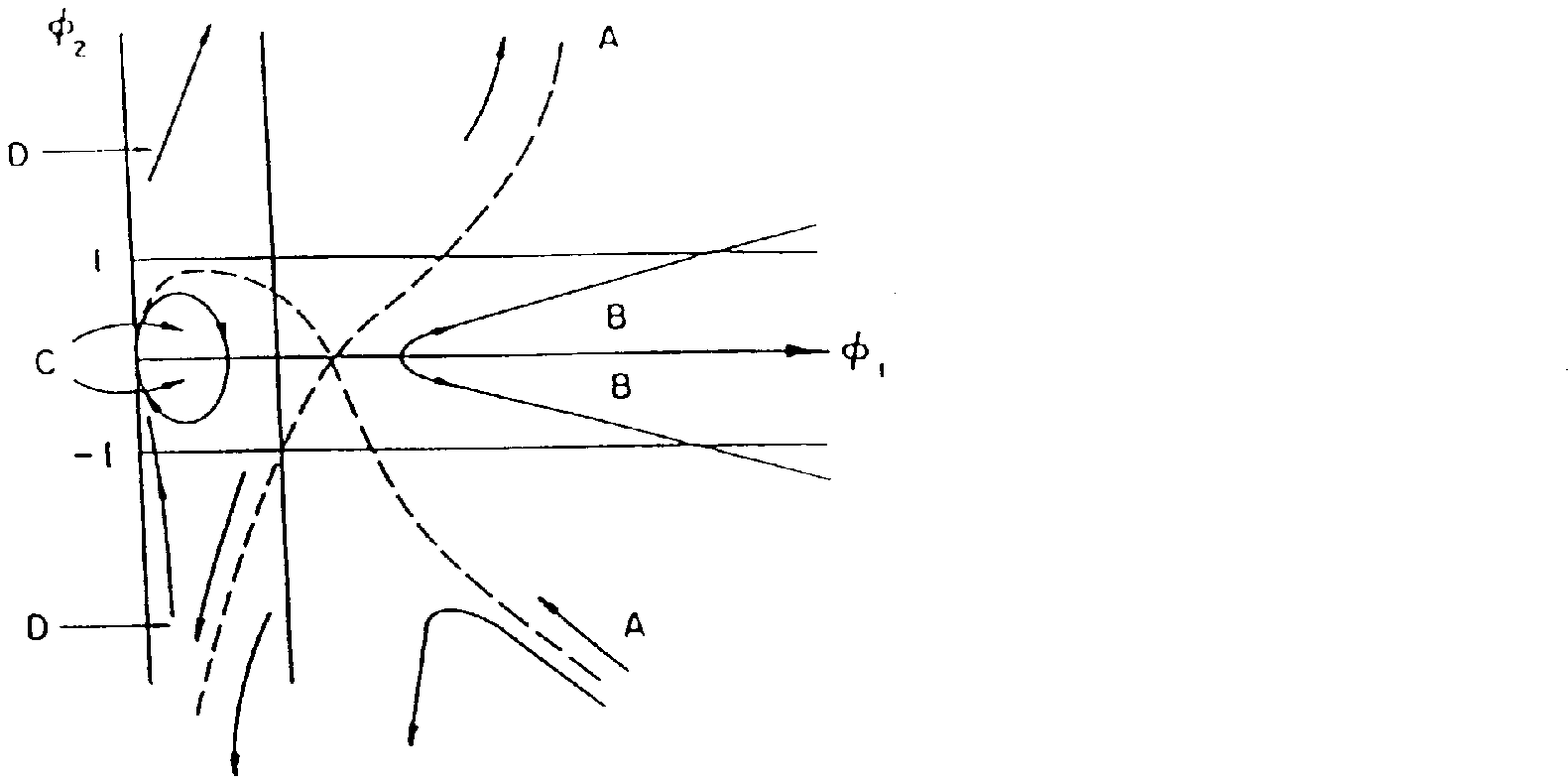}
\caption{
Phase diagram of the homogeneous and isotropic solutions of the nonsingular Universe. The asymptotic regions are labelled by A, B, C and D, flow lines are indicated by arrows.}
\end{center}
\end{figure}

The resulting phase diagram of vacuum solutions is sketched in Fig. 10 
(for numerical results, see \cite{BMS93}).  The point $(\varphi_1, \, 
\varphi_2) = (0,0)$ corresponds to Minkowski space-time $M^4$, the 
regions $|\varphi_2 | \rightarrow \infty$ to de Sitter space.  As 
shown, all solutions either are periodic about $M^4$ or else they 
asymptotically approach de Sitter space.  Hence, all solutions are 
nonsingular.  This conclusion remains unchanged if we add spatial 
curvature to the model.

One of the most interesting properties of our theory is asymptotic 
freedom$^{\cite{BMS93}}$, i.e., the coupling between matter and gravity goes to 
zero at high curvatures.  It is easy to add matter (e.g., dust or 
radiation) to our model by taking the combined action 
\be
S = S_g + S_m \, , 
\ee
where $S_g$ is the gravity action previously discussed, and $S_m$ is 
the usual matter action in an external background space-time metric.

We find$^{\cite{BMS93}}$ that in the asymptotic de Sitter regions, the trajectories of 
the solutions in the $(\varphi_1, \, \varphi_2)$ plane are unchanged 
by adding matter.  This applies, for example, in a phase of de Sitter 
contraction when the matter energy density is increasing exponentially 
but does not affect the metric.  The physical reason for asymptotic 
freedom is obvious: in the asymptotic regions of phase space, the 
space-time curvature approaches its maximal value and thus cannot be 
changed even by adding an arbitrary high matter energy density.

Naturally, the phase space trajectories near $(\varphi_1, \, 
\varphi_2) = (0,0)$ are strongly effected by adding matter.  In 
particular, $M^4$ ceases to be a stable fixed point of the evolution 
equations.
 
\medskip
\centerline{\bf Discussion}
\medskip

We have shown that a class of higher derivative extensions of the 
Einstein theory exist for which many interesting solutions are 
nonsingular.  Our class of models is very special.  Most higher 
derivative theories of gravity have, in fact, much worse singularity 
properties than the Einstein theory.  What is special about our class 
of theories is that they are obtained using a well motivated Lagrange 
multiplier construction which implements the limiting curvature 
hypothesis.  We have shown that \\
\noindent{\rm i)} all homogeneous and isotropic solutions are 
nonsingular$^{\cite{MB92,BMS93}}$\\
\noindent{\rm ii)} the two-dimensional black holes are nonsingular$^{\cite{BMT93}}$\\
\noindent{\rm iii)} nonsingular two-dimensional cosmologies exist$^{\cite{MT94}}$.

By construction, all solutions are de Sitter at high curvature.  Thus, 
the theories automatically have a period of inflation (driven by the 
gravity sector in analogy to Starobinsky inflation$^{\cite{Starob}}$) in the 
early Universe.

A very important property of our theories is asymptotic freedom.  This 
means that the coupling between matter and gravity goes to zero at 
high curvature, and might lead to an automatic suppression mechanism 
for scalar fluctuations.

\subsection{Reheating in Inflationary Cosmology} 

Reheating is an important stage in inflationary cosmology. It determines the state of the Universe after inflation and has consequences for baryogenesis, defect formation, and, as will be shown below, maybe even for the composition
of the dark matter of the Universe.

After slow rolling, the inflaton field begins to oscillate uniformly in space about the true vacuum state. Quantum mechanically, this corresponds to a coherent state of $k = 0$ inflaton particles. Due to interactions of the inflaton with itself and with other fields, the coherent state will decay into quanta of elementary particles. This corresponds to post-inflationary particle production.

Reheating is usually studied using simple scalar field toy models. The one we will adopt here consists of two real scalar fields, the inflaton $\varphi$
with Lagrangian
\be
{\cal L}_o \, = \, {1 \over 2} \partial_\mu \varphi \partial^\mu \varphi - {1 \over 4} \lambda (\varphi^2 - \sigma^2)^2 
\ee
interacting with a massless scalar field $\chi$ representing ordinary matter. The interaction Lagrangian is taken to be
\be
{\cal L}_I \, = \, {1 \over 2} g^2 \varphi^2 \chi^2 \, .
\ee
Self interactions of $\chi$ are neglected. 

By a change of variables
\be
\varphi \, = \, {\tilde \varphi} + \sigma \, ,
\ee
the interaction Lagrangian can be written as
\be \label{intlag}
{\cal L}_I \, = \, g^2 \sigma {\tilde \varphi} \chi^2 + {1 \over 2} g^2 {\tilde \varphi}^2 \chi^2 \, .
\ee
During the phase of coherent oscillations, the field ${\tilde \varphi}$ oscillates with a frequency
\be
\omega \, = \, m_{\varphi} \, = \, \lambda^{1/2} \sigma 
\ee
(neglecting the expansion of the Universe which can be taken into account as in \cite{KLS94,STB95}).

\medskip
\centerline{\bf Elementary Theory of Reheating}
\medskip

According to the elementary theory of reheating (see e.g. \cite{DolLin} and \cite{AFW}), the decay of the inflaton is calculated using first order perturbations theory. According to the Feynman rules, the decay rate $\Gamma_B$ of $\varphi$ (calculated assuming that the cubic coupling term dominates) is
given by
\be
\Gamma_B \, = \, {{g^2 \sigma^2} \over {8 \pi m_{\phi}}} \, .
\ee
The decay leads to a decrease in the amplitude of $\varphi$ (from now on we will drop the tilde sign) which can be approximated by adding an extra damping term to the equation of motion for $\varphi$:
\be
{\ddot \varphi} + 3 H {\dot \varphi} + \Gamma_B {\dot \varphi} \, = \,
- V^\prime(\varphi) \, .
\ee
From the above equation it follows that as long as $H > \Gamma_B$, particle production is negligible. During the phase of coherent oscillation of $\varphi$, the energy density and hence $H$ are decreasing. Thus, eventually $H = \Gamma_B$, and at that point reheating occurs (the remaining energy density in $\varphi$ is very quickly transferred to $\chi$ particles.

The temperature $T_R$ at the completion of reheating can be estimated by computing the temperature of radiation corresponding to the value of $H$ at which $H = \Gamma_B$. From the FRW equations it follows that
\be
T_R \, \sim \, (\Gamma_B m_{pl})^{1/2} \, .
\ee
If we now use the ``naturalness" constraint{\footnote{At one loop order, the cubic interaction term will contribute to $\lambda$ by an amout $\Delta \lambda \sim g^2$. A renormalized value of $\lambda$ smaller than $g^2$ needs to be finely tuned at each order in perturbation theory, which is ``unnatural".}} 
\be
g^2 \, \sim \, \lambda
\ee
in conjunction with the constraint on the value of $\lambda$ from (\ref{fluctconstr}), it follows that for $\sigma < m_{pl}$,
\be
T_R \, < \, 10^{10} {\rm GeV} \, .
\ee
This would imply no GUT baryogenesis, no GUT-scale defect production, and no gravitino problems in supersymmetric models with $m_{3/2} > T_R$, where $m_{3/2}$ is the gravitino mass. As we shall see, these conclusions change radically if we adopt an improved analysis of reheating.

\medskip
\centerline{\bf Modern Theory of Reheating}
\medskip

However, as was first realized in \cite{TB90}, the above analysis misses an essential point. To see this, we focus on the equation of motion for the matter field $\chi$ coupled to the inflaton $\varphi$ via the interaction Lagrangian ${\cal L}_I$ of (\ref{intlag}). Taking into account for the moment only the cubic interaction term, the equation of motion becomes
\be
{\ddot \chi} + 3H{\dot \chi} - \bigl(({{\nabla} \over a})^2 - m_{\chi}^2 - 2g^2\sigma\varphi \bigr)\chi \, = \, 0 \, .
\ee
Since the equation is linear in $\chi$, the equations for the Fourier modes $\chi_k$ decouple:
\be \label{reseq}
{\ddot \chi_k} + 3H{\dot \chi_k} + (k_p^2 + m_{\chi}^2 + 2g^2\sigma\varphi)\chi_k \, = \, 0 ,
\ee
where $k_p$ is the time-dependent physical wavenumber. 

Let us for the moment neglect the expansion of the Universe. In this case, the friction term in (\ref{reseq}) drops out and $k_p$ is time-independent, and Equation (\ref{reseq}) becomes a harmonic oscillator equation with a time-dependent mass determined by the dynamics of $\varphi$. In the reheating phase, $\varphi$ is undergoing oscillations. Thus, the mass in (\ref{reseq}) is varying periodically. In the mathematics literature, this equation is called the Mathieu equation. It is well known that there is an instability. In physics, the effect is known as {\bf parametric resonance} (see e.g. \cite{parres}). At frequencies $\omega_n$ corresponding to half integer multiples of the frequency $\omega$ of
the variation of the mass, i.e.
\be
\omega_k^2 = k_p^2 + m_{\chi}^2 \, = \, ({n \over 2} \omega)^2 \,\,\,\,\,\,\, n = 1, 2, ... ,
\ee
there are instability bands with widths $\Delta \omega_n$. For values of $\omega_k$ within the instability band, the value of $\chi_k$ increases exponentially:
\be
\chi_k \, \sim \, e^{\mu t} \,\,\,\, {\rm with} \,\,\, \mu \sim {{g^2 \sigma \varphi_0} \over {\omega}} \, ,
\ee
with $\varphi_0$ being the amplitude of the oscillation of $\varphi$. Since the widths of the instability bands decrease as a power of the (small) coupling constant $g^2$ with increasing $n$, for practical purposes only the lowest instability band is important. Its width is
\be
\Delta \omega_k \, \sim \, g \sigma^{1/2} \varphi_0^{1/2} \, .
\ee
Note, in particular, that there is no ultraviolet divergence in computing the total energy transfer from the $\varphi$ to the $\chi$ field due to parametric resonance.

It is easy to include the effects of the expansion of the Universe (see e.g. \cite{TB90,KLS94,STB95}). The main effect is that the value of $\omega_k$ becomes time-dependent. Thus, a mode slowly enters and leaves the resonance bands. As a consequence, any mode lies in the resonance band for only a finite time. This implies that the calculation of energy transfer is perfectly well-behaved. No infinite time divergences arise.

It is now possible to estimate the rate of energy transfer, whose order of magnitude is given by the phase space volume of the lowest instability band multiplied by the rate of growth of the mode function $\chi_k$. Using as an initial condition for $\chi_k$ the value $\chi_k \sim H$ given by the magnitude of the expected quantum fluctuations, we obtain
\be \label{entransf}
{\dot \rho} \, \sim \, \mu ({\omega \over 2})^2 \Delta\omega_k H e^{\mu t} \, .
\ee

From (\ref{entransf}) it follows that provided that the condition
\be \label{rescond}
\mu \Delta t \, >> 1
\ee
is satisfied, where $\Delta t < H^{-1}$ is the time a mode spends in the instability band, then the energy transfer will procede fast on the time scale
of the expansion of the Universe. In this case, there will be explosive particle production, and the energy density in matter at the end of reheating will be given by the energy density at the end of inflation.  

The above is a summary of the main physics of the modern theory of reheating.
The actual analysis can be refined in many ways (see e.g. \cite{KLS94,STB95}).
First of all, it is easy to take the expansion of the Universe into account
explicitly (by means of a transformation of variables), to employ an exact solution of the background model and to reduce the mode equation for $\chi_k$ to a Hill equation, an equation similar to the Mathieu equation which also admits exponential instabilities.

The next improvement consists of treating the $\chi$ field quantum mechanically (keeping $\varphi$ as a classical background field). At this point, the techniques of quantum field theory in a curved background can be applied. There is no need to impose artificial classical initial conditions for $\chi_k$. Instead, we may assume that $\chi$ starts in its initial vacuum state (excitation of an initial thermal state has been studied in \cite{Yoshimura2}), and the Bogoliubov mode mixing technique (see e.g. \cite{Birrell}) can be used to compute the number of particles at late times.

Using this improved analysis, we recover the result (\ref{entransf}). Thus, provided that the condition (\ref{rescond}) is satisfied, reheating will be explosive. Working out the time $\Delta t$ that a mode remains in the instability band for our model, expressing $H$ in terms of $\varphi_0$ and $m_{pl}$, and $\omega$ in terms of $\sigma$, and using the naturalness relation $g^2 \sim \lambda$, the condition for explosive particle production becomes
\be \label{rescond2}
{{\varphi_0 m_{pl}} \over {\sigma^2}} \, >> \, 1 \, ,
\ee
which is satisfied for all chaotic inflation models with $\sigma < m_{pl}$ (recall that slow rolling ends when $\varphi \sim m_{pl}$ and that therefore the initial amplitude $\varphi_0$ of oscillation is of the order $m_{pl}$).

We conclude that rather generically, reheating in chaotic inflation models will be explosive. This implies that the energy density after reheating will be approximately equal to the energy density at the end of the slow rolling period. Therefore, as suggested in \cite{KLS96,Tkachev} and \cite{KLR96}, respectively, GUT scale defects may be produced after reheating and GUT-scale baryogenesis scenarios may be realized, provided that the GUT energy scale is lower than
the energy scale at the end of slow rolling.

Note, however, that the state of $\chi$ after parametric resonance is {\bf not} a thermal state. The spectrum consists of high peaks in distinct wave bands. An important question which remains to be studied is how this state thermalizes.
For some interesting work on this issue see \cite{therm}. As emphasized in \cite{KLS96} and \cite{Tkachev}, the large peaks in the spectrum may lead to symmetry restoration and to the efficient production of topological defects (for a differing view on this issue see \cite{AC96,Boyan2}). Since the state after explosive particle production is not a thermal state, it is useful to follow
\cite{KLS94} and call this process ``preheating" instead of reheating.

A futher interesting conjecture which emerges from the parametric resonance analysis of preheating$^{\cite{KLS94,STB95}}$ is that the dark matter in the Universe may consist of remnant coherent oscillations of the inflaton field. In fact, it can easily be checked from (\ref{rescond2}) that the condition for efficient transfer of energy eventually breaks down when $\varphi_0$ has decreased to a sufficiently small value. For the model considered here, an order of magnitude calculation shows that the remnant oscillations may well contribute significantly to the present value of $\Omega$.

Note that the details of the analysis of preheating are quite model-dependent. In fact$^{\cite{KLS94}}$, in many models one does not get the kind of ``narrow-band" resonance discussed here, but ``wide-band" resonance. In this case, the energy transfer is even more efficient.

There has recently been a lot of work on various aspects of reheating (see e.g. \cite{Yoshimura1,Boyan1,Kaiser,ALR96} for different approaches). Many important questions, e.g. concerning thermalization and back-reaction effects during and after preheating (or parametric resonance) remain to be fully analyzed.

\subsection{Summary}

The inflationary Universe is an attractive scenario for early Universe cosmology. It can resolve some of the problems of standard cosmology, and in addition gives rise to a predictive theory of structure formation (see e.g. \cite{Liddle} for a recent review).

However, important unsolved problems of principle remain. Rather generically, the predicted amplitude of perturbations is too large (the spectral shape, however, is in quite good agreement with the observations). The present realizations of inflation based on scalar field also make the cosmological constant problem more accute. In addition, there are no convincing particle-physics based realizations of inflation. Many models of inflation resort to introducing a new matter sector. It is important to search for a better connection between modern particle physics / field theory and inflation.
String cosmology and dilaton gravity (see e.g. the recent reviews in \cite{stringcos}) may provide an interesting new approach to the unification of inflation and fundamental physics.

Recently, there has been much progress in the understanding of the energy transfer at the end of inflation between the inflaton field and matter. It appears that resonance phenomena such as parametric resonance play a crucial role. These new reheating scenarios lead to a high reheating temperature,
although much more work remains to be done before one can reach a final conclusion on this issue.

\section{Classical and Quantum Theory of Cosmological Perturbations}

In inflationary Universe and topological defect models of structure formation, small amplitude seed perturbations are predicted to arise due to particle physics effects in the very early Universe.  They then grow by gravitational
instability to produce the cosmological structures we observe today.  In order
to be able to make the connection between particle physics and observations, it
is important to understand the gravitational evolution of fluctuations.  This
section will introduce the basic concepts of this topic.   

As is evident from Figure 5 and from the discussion of inflation in the previous section, general relativity and quantum mechanics both play a fundamental role in the theory of perturbations. In inflationary Universe models, quantum effects seed the fluctuations, and thus a quantum analysis of the generation of fluctuations is essential. However, since the fluctuations are small, a linearized analysis is sufficient. Since the scales on which we are interested in following the fluctuations are larger than the Hubble radius for a long time interval, Newtonian gravity is obviously inadequate to treat these perturbations, and general relativistic effects become essential.

In this section, we will first introduce some basic notation, then discuss the Newtonian theory of linear fluctuations before turning to the full relativistic analysis.

\subsection{Basic Issues}

In this article we only discuss theories in which structures grow by
gravitational accretion.  The basic mechanism is easy to understand.
Consider first a flat space-time background.  A density perturbation with
$\delta \rho > 0$ leads to an excess gravitational attractive
force $F$ acting on the surrounding matter.  This force is proportional to
$\delta \rho$, and will hence lead to exponential growth of the perturbation
since
\be \label{intu1}
\delta \ddot \rho \sim F \sim \delta \rho \Rightarrow \delta \rho \sim \exp
(\alpha t) 
\ee
with a constant $\alpha$ which is proportional to Newton's constant $G$.

In an expanding background space-time, the acceleration is damped by the
expansion.  If $r (t)$ is the physical distance of a test particle from the
perturbation, then on a scale $r$
\be \label{intu2}
\delta \ddot \rho \sim F \sim \, {\delta \rho\over{r^2 (t)}} \, , 
\ee
which results in power-law increase of $\delta \rho$.  The goal of this
subsection is to discuss the growth rates of inhomogeneities in more detail
(see e.g. \cite{Efstath,Padmanabhan} for modern reviews).

Because of our assumption that all perturbations start out with  a small
amplitude, we can linearize the equations for gravitational fluctuations.  The
analysis is then greatly simplified by going to momentum space in which all
modes $\delta (\underline{k})$ decouple.  We expand the fractional density
contrast $\delta (\underline{x})$ as follows:
\be \label{Fourier}
\delta (\underline{x}) = {{\delta \rho (\underline{x})} \over \rho} = (2 \pi)^{-3/2} V^{1/2}
\int d^3 k \> e^{i \underline{k} \cdot \underline{x}} \delta (\underline{k}) \,
, 
\ee
where $V$ is a cutoff volume which disappears from all physical observables.

The ``power spectrum" $P(k)$ is defined by
\be \label{powersp}
P (k) = < |\delta (k) |^2 > \, , 
\ee
where the braces denote an ensemble average (in most structure formation
models, the generation of perturbations is a stochastic process, and hence
observables can only be calculated by averaging over the ensemble.  For
observations, the braces can be viewed as an angular average).

The physical measure of mass fluctuations on a length scale $\lambda$ is the
r.m.s. mass fluctuation $\delta M/M (\lambda)$ on this scale.  It is determined
by the power spectrum in the following way.  We pick a center $\underline{x}_0$
of a sphere $B_\lambda (\underline{x}_0)$ of radius $\lambda$ and calculate
\be
\big| {\delta M\over M} \big|^2 \, (\underline{x}_0 , \, \lambda) = \big|
\int\limits_{B_\lambda (\underline{x}_0)} d^3 x \delta (\underline{x}) \,
{1\over{V (B_\lambda)}} \big|^2 \, , 
\ee
where $V (B_\lambda)$ is the volume of the sphere.  Inserting the Fourier
decomposition (\ref{Fourier}) and taking the average value of this quantity over all $\underline{x}_0$ yields
\be
< \left( {\delta M\over M} \right)^2 (\lambda) > = \int d^3 k W_k (\lambda) |
\delta (\underline{k}) |^2 
\ee
with a window function $W_k (\lambda)$ with the following properties
\be
W_k (\lambda) \cases{\simeq 1 & $k < k_\lambda = 2 \pi / \lambda$ \cr
\simeq 0 & $k > k_\lambda$ .\cr} 
\ee
Therefore the r.m.s. mass perturbation on a scale $\lambda$ becomes
\be \label{masspow}
< \big| {\delta M\over M} (\lambda) \big|^2 > \sim k_{\lambda}^3 P(k_{\lambda})
\, . 
\ee

If $P (k) \sim k^n$ then $n$ is called the index of the power spectrum.  For $n = 1$ we get the
so-called Harrison-Zel'dovich scale invariant spectrum$^{\cite{HZ}}$.

Both inflationary Universe and topological defect models of structure formation
predict a roughly scale invariant spectrum.  The distinguishing feature of this
spectrum is that the r.m.s. mass perturbations are independent of the scale $k$
when measured at the time $t_H (k)$ when the associated wavelength is equal to
the Hubble radius, i.e., when the scale ``enters" the Hubble radius.  Let us
derive this fact for the scales entering during the matter dominated epoch.
The time $t_H (k)$ is determined by
\be \label{Hubble}
k^{-1} a (t_H (k)) = t_H (k) 
\ee
which leads to $t_H (k) \sim k^{-3}$.
According to the linear theory of cosmological perturbations discussed in the
following subsection, the mass fluctuations increase as $a(t)$ for $t >
t_{eq}$.  Hence
\be \label{growth}
{\delta M\over M} (k, t_H (k)) = \left( {t_H (k)\over t} \right)^{2/3} \,
{\delta M\over M} \, (k, t) \sim {\rm const} \, , 
\ee
since the first factor scales (from (\ref{Hubble}) as $k^{-2}$ and -- using (\ref{masspow}) and inserting
$n=1$ -- the second as $k^2$.

\subsection{Newtonian Theory}

The Newtonian theory of cosmological perturbations is an approximate analysis
which is valid on wavelengths $\lambda$ much smaller than the Hubble radius $t$
and for negligible pressure $p$, i.e., $p \ll  \rho$.  It is based on expanding
the hydrodynamical equations about a homogeneous background solution.

The starting points are the continuity, Euler and Poisson equations
\be \label{Newteq1}
\dot \rho + \underline{\nabla} (\rho \underline{v}) = 0 
\ee
\be \label{Newteq2}
\underline{\dot v} + (\underline{v} \cdot \underline{\nabla}) \underline{v} = -
\underline{\nabla} \phi - {1\over \rho} \, \underline{\nabla} p 
\ee
\be  \label{Newteq3}
\nabla^2  \phi = 4 \pi G \rho 
\ee
for a fluid with energy density $\rho$, pressure $p$, velocity $\underline{v}$
and Newtonian gravitational potential $\phi$, written in terms of physical
coordinates $(t, \, \underline{r})$.

The transition to an expanding space is made by introducing comoving
coordinates $\underline{x}$ and peculiar velocity $\underline{u} =
\underline{\dot {x}}$:
\be
\underline{r} = a (t) \underline{x} 
\ee
\be \label{vel}
\underline{v} = \dot a (t) \underline{x} + a (t) \underline{u} \, . 
\ee
The first term on the right hand side of (\ref{vel}) is the expansion velocity.

The perturbation equations are obtained by linearizing Equations (\ref{Newteq1} - \ref{Newteq3})
about a homogeneous background solution $\rho = \bar \rho (t) , \> p = 0$ and
$\underline{u} = 0$.  The linearization ansatz can be written
\be
\rho (\underline{x}, t) = \bar \rho (t) (1 + \delta (\underline{x}, t) ) \,\,\,
|\delta| << 1 .
\ee
If we consider adiabatic perturbations (no entropy density variations), then
after some algebra the linearized equations become
\be \label{Newteq4}
\dot \delta + \nabla \cdot \underline{u} = 0 \, , 
\ee
\be \label{Newteq5}
\underline{\dot {u}} + 2 {\dot a\over a} \underline{u} = - a^2 (\nabla \delta
\phi + c^2_s \nabla \delta) 
\ee
and
\be \label{Newteq6}
\nabla^2 \delta \phi = 4 \pi G \bar \rho a^2 \delta \, , 
\ee
with the speed of sound $c_s$ given by
\be
c^2_s = {\partial p\over{\partial \rho}} \, . 
\ee
The two first order equations (\ref{Newteq4}) and (\ref{Newteq5}) can be combined to yield a
single second order differential equation for $\delta$.  With the help of
(\ref{Newteq6}) this equation reads
\be
\ddot \delta + 2 H \dot \delta -  4 \pi G \bar \rho \delta - {c^2_s\over a^2}
\nabla^2 \delta = 0 
\ee
which in momentum space becomes
\be \label{momeq}
\ddot \delta_{\underline{k}} + 2 H \dot \delta_{\underline{k}} + \left( {c^2_s
k^2 \over a^2} - 4 \pi G \bar \rho \right) \delta_{\underline{k}} = 0 \, .
\ee
Here, $H(t)$ as usual denotes the expansion rate, and $\delta_{\underline{k}}$
stands for $\delta (\underline{k})$.

Already a quick look at Equation (\ref{momeq}) reveals the presence of a distinguished
scale for cosmological perturbations, the Jeans length
\be
\lambda_J = {2 \pi\over k_J} 
\ee
with
\be \label{Jeans}
k^2_J = \left( {k\over a} \right)^2 = {4 \pi G \bar \rho\over c^2_s} \, .
\ee
On length scales larger than $\lambda_J$, the spatial gradient term is
negligible, and the term linear in $\delta$ in (\ref{momeq}) acts like a negative mass
square quadratic potential with damping due to the expansion of the Universe,
in agreement with the intuitive analysis leading to (ref{intu1}) and (\ref{intu2}).  On length
scales smaller than $\lambda_J$, however, (\ref{momeq}) becomes a damped harmonic oscillator equation and perturbations on these scales decay.

For $t > t_{eq}$ and for $\lambda \gg \lambda_J$, Equation (\ref{momeq}) becomes
\be
\ddot \delta_k + {4\over{3t}} \, \dot \delta_k - {2\over{3t^2}} \, \delta_k = 0
\ee
and has the general solution
\be
\delta_k (t) = c_1 t^{2/3} + c_2 t^{-1} \, . 
\ee
This demonstrates that for $t > t_{eq}$ and $\lambda \gg \lambda_J$, the
dominant mode of perturbations increases as $a(t)$, a result we already used in
the previous subsection (see (\ref{growth})).

For $\lambda \ll \lambda_J$ and $t > t_{eq}$, Equation (\ref{momeq}) becomes
\be
\ddot \delta_k + 2 H \dot \delta_k + c_s^2 \left({k\over a} \right)^2 \delta_k
= 0 \, , 
\ee
and has solutions corresponding to damped oscillations:
\be
\delta_k (t) \sim a^{-1/2} (t) \exp \{ \pm i c_s k \int dt^\prime a
(t^\prime)^{-1} \} \, . 
\ee

As an important application of the Newtonian theory of cosmological
perturbations, let us compare sub-horizon scale fluctuations in a
baryon-dominated Universe $(\Omega = \Omega_B = 1)$ and in a CDM-dominated
Universe with $\Omega_{CDM} = 0.9$ and $\Omega = 1$.  We consider scales which
enter the Hubble radius at about $t_{eq}$.

In the initial time interval $t_{eq} < t < t_{rec}$, the baryons are coupled to
the photons.  Hence, the baryonic fluid has a large pressure $p_B$
\be
p_B \simeq p_r = {1\over 3} \, \rho_r \, , 
\ee
and therefore the speed of sound is relativistic
\be
c_s \simeq \left( {p_r\over \rho_m} \right)^{1/2} = {1\over{\sqrt{3}}} \,
\left({\rho_r\over \rho_m} \right)^{1/2} \, . 
\ee
The value of $c_s$ slowly decreases in this time interval, attaining a value of
about $1/10$ at $t_{rec}$.  From (\ref{Jeans}) it follows that the Jeans mass $M_J$,
the mass inside a sphere of radius $\lambda_J$, increases until $t_{rec}$ when
it reaches its maximal value $M_J^{max}$
\be
M_J^{max} = M_J (t_{rec}) = {4 \pi\over 3} \, \lambda_J (t_{rec})^3 \bar \rho
(t_{rec}) \sim 10^{17} (\Omega h^2)^{-1/2} M_{\odot} \, . 
\ee

At the time of recombination, the baryons decouple from the radiation fluid.
Hence, the baryon pressure $p_B$ drops abruptly, as does the Jeans length (see
(\ref{Jeans})).  The remaining pressure $p_B$ is determined by the temperature and
thus continues to decrease as $t$ increases.  It can be shown that the Jeans
mass continues to decrease after $t_{rec}$, starting from a value
\be
M^-_J (t_{rec}) \sim 10^6 (\Omega h^2)^{-1/2} \, M_\odot 
\ee
(where the superscript ``$-$" indicates the mass immediately after $t_{eq}$.

In contrast, CDM has negligible pressure throughout the period $t > t_{eq}$ and
hence experiences no Jeans damping.  A CDM perturbation which enters the Hubble
radius at $t_{eq}$ with amplitude $\delta_i$ has an amplitude at $t_{rec}$
given by
\be
\delta^{CDM}_k (t_{rec}) \simeq \, {a (t_{rec})\over{a (t_{eq})}} \, \delta_i
\, , 
\ee
whereas a perturbation with the same scale and initial amplitude in a
baryon-dominated Universe is damped
\be
\delta_k^{BDM} (t_{rec}) \simeq \, \left({a (t_{rec})\over{a (t_{eq})}}
\right)^{-1/2} \, \delta_i \, . 
\ee
In order for the perturbations to have the same amplitude today, the initial
size of the inhomogeneity must be much larger in a BDM-dominated Universe than
in a CDM-dominated one:
\be
\delta^{BDM}_k (t_{eq}) \simeq \left( {z (t_{eq})\over{z (t_{rec})}}
\right)^{3/2} \delta_k^{CDM} \, (t_{eq}) \, . 
\ee
For $\Omega = 1$ and $h = 1/2$ the enhancement factor is about 30.

In a CDM-dominated Universe the baryons experience Jeans damping, but after
$t_{rec}$ the baryons quickly fall into the potential wells created by the CDM
perturbations, and hence the baryon perturbations are proportional to the CDM
inhomogeneities.  

The above considerations, coupled with information about CMB anisotropies, can
be used to rule out a model with $\Omega = \Omega_B = 1$.  The argument goes as
follows.  For adiabatic fluctuations, the amplitude of CMB
anisotropies on an angular scale $\vartheta$ is determined by the value of
$\delta \rho/\rho$ (strictly speaking, the relativistic potential $\Phi$ to be discussed in the following subsection) on the corresponding length scale $\lambda (\vartheta)$ at
$t_{eq}$:
\be
{\delta T\over T} (\vartheta)  = {1\over 3} \, {{\delta \rho} \over \rho} (
\lambda (\vartheta), \, t_{eq} ) \, . 
\ee
 On scales of clusters we know that (for $\Omega = 1$ and $h = 1/2$)
\be
\left({{\delta \rho} \over \rho} \right)_{CDM} \, (\lambda (\vartheta), \, t_{eq})
\simeq z (t_{eq})^{-1} \simeq 10^{-4} \, , 
\ee
using the fact that today on cluster scales $\delta \rho/\rho \simeq 1$.  The
bounds on $\delta T/ T$ on small angular scales are
\be
{\delta T\over T}(\vartheta)  << 10^{-4} \, , 
\ee
consistent with the predictions for a CDM model, but inconsistent with those of
a $\Omega = \Omega_B = 1$ model, according to which we would expect
anisotropies of the order of $10^{-3}$.  This is yet another argument in
support of the existence of nonbaryonic dark matter.

To conclude this subsection, let us briefly discuss two further aspects related to Newtonian perturbations. The first concerns matter inhomogeneities
during the radiation-dominated epoch.  We consider matter fluctuations with
$c_s = 0$ in a smooth relativistic background.  In this case, Equation (\ref{momeq})
becomes
\be \label{meszaros1}
\ddot \delta_k + 2 H \dot \delta_k - 4 \pi G \bar \rho_m \delta_k = 0 \, ,
\ee
where $\bar \rho_m$ denotes the average matter energy density.  The Hubble
expansion parameter obeys
\be \label{meszaros2}
H^2 = {8 \pi G\over 3} (\bar \rho_m + \bar \rho_r) \, , 
\ee
with $\bar \rho_r$ the background radiation energy density.  For $t < t_{eq}$,
$\bar \rho_m$ is negligible in both (\ref{meszaros1}) and (\ref{meszaros2}), and (\ref{meszaros1}) has the general solution
\be
\delta_k (t) = c_1 \log t + c_2 \, . 
\ee
In particular, this result implies that CDM perturbations which enter the
Hubble radius before $t_{eq}$ have an amplitude which grows only
logarithmically in time until $t_{eq}$. 

Finally, we consider hot dark matter (HDM) fluctuations. Whereas CDM particles are cold, i.e. their peculiar velocity is negligible for all times relevant for structure formation, HDM particles have relativistic velocities at $t_{eq}$, i.e. $v(t_{eq}) \sim 1$. The prime candidate for HDM is a $25h_{50}^2$eV tau neutrino.

The new aspect of HDM is related to neutrino free
streaming$^{\cite{freestr}}$. Because of the large velocity of the dark matter particles, pure dark matter inhomogeneities are washed out on all scales below the neutrino free streaming length  $\lambda^c_j (t)$,
\be \label{Jeansscale}
\lambda^c_j (t) \sim v (t) z (t) t \, ,  
\ee
which is the comoving distance the particles move in one Hubble expansion time.
Since the neutrino velocity $v(t)$ and the redshift $z(t)$ both scale as
$a(t)^{-1}$,  the free streaming length decreases as
\be \label{Jeansdecay}
\lambda^c_j (t) \sim t^{-1/3} 
\ee
after $t_{eq}$ (before $t_{eq}$ the radiation pressure dominates).  

Hence, in an inflationary HDM model in which the fluctuations are dark matter inhomogeneities,  all perturbations on scales $\lambda$ smaller than
the maximal value of $\lambda^c_j (t)$ are erased.  The critical scale
$\lambda^{\rm max}_j$ is given by the value of $\lambda^c_j (t)$ at the time
when the neutrinos become non-relativistic, which is in turn determined by the
neutrino mass $m_{\nu}$.  The result is
\be \label{freestream}
\lambda^{\rm max}_j \simeq 30 \, {\rm Mpc} \, \left({{m_{\nu}} \over {25 {\rm
eV}}} \right)^{-2} \, , 
\ee
a scale much larger than the mean separation of galaxies and clusters.  Since
we observe galaxies outside of large-scale structures, this model is in blatant
disagreement with observations. However, theories in which the primordial perturbations are nonadiabatic long-lived seeds (e.g. cosmic strings), may well be viable if the dark matter is hot. As we shall see in Section 4, the cosmic string model in fact works well for hot dark matter.

\subsection{Relativistic Theory: Classical Analysis}

On scales larger than the Hubble radius $(\lambda > t)$ the Newtonian theory of
cosmological perturbations obviously is inapplicable, and a general
relativistic analysis is needed.  On these scales, matter is essentially frozen
in comoving coordinates.  However, space-time fluctuations can still increase
in amplitude.

In principle, it is straightforward to work out the general relativistic theory
of linear fluctuations$^{\cite{Lifshitz}}$.  We linearize the Einstein  equations
\be
G_{\mu\nu} = 8 \pi G T_{\mu\nu} 
\ee
(where $G_{\mu\nu}$ is the Einstein tensor associated with the space-time
metric $g_{\mu\nu}$, and $T_{\mu\nu}$ is the energy-momentum tensor of matter)
about an expanding FRW background $(g^{(0)}_{\mu\nu} ,\, \varphi^{(0)})$:
\beq
g_{\mu\nu} (\underline{x}, t) & = & g^{(0)}_{\mu\nu} (t) + h_{\mu\nu}
(\underline{x}, t) \\
\varphi (\underline{x}, t) & = & \varphi^{(0)} (t) + \delta \varphi
(\underline{x}, t) \,  
\eeq
and pick out the terms linear in $h_{\mu\nu}$ and $\delta \varphi$ to obtain
\be \label{linein}
\delta G_{\mu\nu} \> = \> 8 \pi G \delta T_{\mu\nu} \, . 
\ee
In the above, $h_{\mu\nu}$ is the perturbation in the metric and $\delta
\varphi$ is the fluctuation of the matter field $\varphi$.  We have denoted all
matter fields collectively by $\varphi$.

In practice, there are many complications which make this analysis highly
nontrivial.  The first problem is ``gauge invariance"$^{\cite{PressVish}}$  Imagine starting
with a homogeneous FRW cosmology and introducing new coordinates which mix
$\underline{x}$ and $t$.  In terms of the new coordinates, the metric now looks
inhomogeneous.  The inhomogeneous piece of the metric, however, must be a pure
coordinate (or "gauge") artefact.  Thus, when analyzing relativistic
perturbations, care must be taken to factor out effects due to coordinate
transformations.

\begin{figure}
\begin{center}
\leavevmode
\epsfxsize=11cm \epsfbox{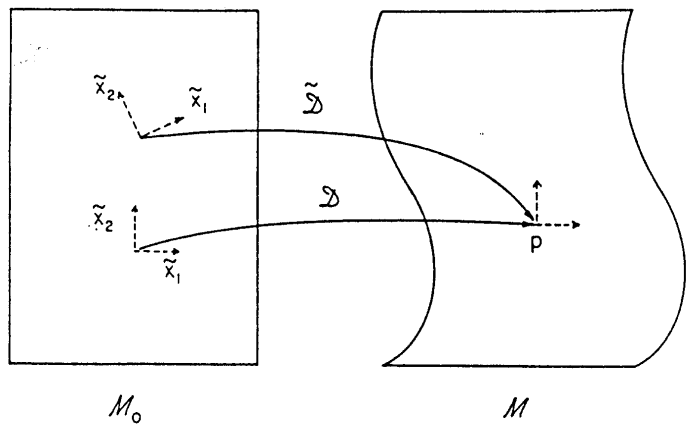}
\caption{
Sketch of how two choices of the mapping from the
background space-time manifold ${\cal M}_0$ to the physical manifold ${\cal M}$
induce two different coordinate systems on ${\cal M}$.}
\end{center}
\end{figure}  

The issue of gauge dependence is illustrated in Fig. 11.  A coordinate system
on the physical inhomogeneous space-time manifold ${\cal M}$ can be viewed as a
mapping ${\cal D}$ of an unperturbed space-time ${\cal M}_0$ into ${\cal M}$.
A physical quantity $Q$ is a geometrical function defined on ${\cal M}$.  There
is a corresponding physical quantity $^{(0)}Q$ defined on ${\cal M}_0$.  In the
coordinate system given by ${\cal D}$, the perturbation $\delta Q$ of $Q$ at
the space-time point $p \, \epsilon \, {\cal M}$ is
\be
\delta Q (p) = Q (p) - \,^{(0)}Q \, (D^{-1} (p) ) \, . 
\ee
However, in a second coordinate system $\tilde {\cal D}$ the perturbation is
given by
\be
\delta \tilde Q (p) = Q (p) - \,^{(0)}Q (\tilde {\cal D}^{-1} (p) ) \, .
\ee
The difference
\be
\Delta Q (p) = \delta Q (p) - \delta \tilde Q (p) 
\ee
is obviously a gauge artefact and carries no physical meaning.

There are various methods of dealing with gauge artefacts.  The simplest and
most physical approach is to focus on gauge invariant variables, i.e.,
combinations of the metric and matter perturbations which are invariant under
linear coordinate transformations.

The gauge invariant theory of cosmological perturbations is in principle
straightforward, although technically rather tedious. In the following I will
summarize the main steps and refer the reader to \cite{MFB92} for the details and further references (see also \cite{MFBrev} for a pedagogical introduction and \cite{Bardeen,BKP83,KoSa84,Durrer,Lyth,Hwang,EllisBruni,Salopek} for other approaches).

We consider perturbations about a spatially flat Friedmann-Robertson-Walker
metric
\be
ds^2 = a^2 (\eta) (d\eta^2 - d \underline{x}^2) 
\ee
where $\eta$ is conformal time (related to cosmic time $t$ by $a(\eta)  d \eta
= dt$).  A scalar metric perturbation (see \cite{Stewart} for a precise definition)
can be written in terms of four free functions of space and time:
\be
\delta g_{\mu\nu} = a^2 (\eta) \pmatrix{2 \phi & -B_{,i} \cr
-B_{,i} & 2 (\psi \delta_{ij} + E_{,ij}) \cr} \, . 
\ee
Scalar metric perturbations are the only perturbations which couple to energy
density and pressure.

The next step is to consider infinitesimal coordinate transformations
\be
x^{\mu^\prime} = x^\mu + \xi^\mu 
\ee
which preserve the scalar nature of $\delta g_{\mu\nu}$, and to calculate the
induced transformations of $\phi, \psi, B$ and $E$.  Then we find invariant
combinations to linear order.  (Note that there are in general no combinations
which are invariant to all orders$^{\cite{SteWa}}$.)  After some algebra, it follows
that
\beq
\Phi & = & \phi + a^{-1} [(B - E^\prime) a]^\prime \\
\Psi & = & \psi - {a^\prime\over a} \, (B - E^\prime)  
\eeq
are two invariant combinations.  In the above, a prime denotes differentiation
with respect to $\eta$.

There are various methods to derive the equations of motion for gauge invariant
variables.  Perhaps the simplest way$^{\cite{MFB92}}$ is to consider the linearized
Einstein equations (\ref{linein}) and to write them out in the longitudinal gauge defined by
\be
B = E = 0 
\ee
and in which $\Phi = \phi$ and $\Psi = \psi$, to directly obtain gauge
invariant equations.

For several types of matter, in particular for scalar field matter, the
perturbation of $T_{\mu \nu}$ has the special property
\be \label{diagonal}
\delta T^i_j \sim \delta^i_j 
\ee
which imples $\Phi = \Psi$.  Hence, the scalar-type cosmological perturbations
can in this case be described by a single gauge invariant variable.  The
equation of motion takes the form$^{\cite{BST,BK84,Lyth,RBrev}}$
\be \label{conserv}
\dot \xi = O \left({k\over{aH}} \right)^2 H \xi 
\ee
where
\be
\xi = {2\over 3} \, {H^{-1} \dot \Phi + \Phi\over{1 + w}} + \Phi \, . 
\ee

The variable $w = p/ \rho$ (with $p$ and $\rho$ background pressure and energy
density respectively) is a measure of the background equation of state.  In
particular, on scales larger than the Hubble radius, the right hand side of
(\ref{conserv}) is negligible, and hence $\xi$ is constant.

The result that $\dot \xi = 0$ is a very powerful one.  Let us first imagine
that the equation of state of matter is constant, {\it i.e.}, $w = {\rm
const}$.  In this case, $\dot \xi = 0$ implies
\be \label{conserv2}
\Phi (t) = {\rm const} \, , 
\ee
{\it i.e.}, this gauge invariant measure of perturbations remains constant
outside the Hubble radius.

Next, consider the evolution of $\Phi$ during a phase transition from an
initial phase with $w = w_i$  to a phase with $w = w_f$.  Long before and after
 the transition, $\Phi$ is constant because of (\ref{conserv2}), and hence $\dot \xi = 0$
becomes
\be
{\Phi\over{1 + w}} + \Phi = {\rm const} \, , 
\ee

In order to make contact with matter perturbations and Newtonian intuition, it
is important to remark that,  as a consequence of the Einstein constraint
equations, at Hubble radius crossing $\Phi$ is a measure of the fractional
density fluctuations:
\be
\Phi (k, t_H (k) ) \sim {\delta \rho\over \rho} \, ( k , \, t_H (k) ) \, .
\ee
(Note that the latter quantity is approximately gauge invariant on scales
smaller than the Hubble radius).

\subsection{Relativistic Theory: Quantum Analysis}

The question of the origin of classical density perturbations from quantum
fluctuations in the de Sitter phase of an inflationary Universe is a rather
subtle issue.  Starting from a homogeneous quantum state ({\it e.g.}, the
vacuum state in the FRW coordinate frame at time $t_i$, the beginning of
inflation), a naive semiclassical anaylsis would predict the absence of
fluctuations since $< \psi | T_{\mu\nu} (x) | \psi >$ is independent of space.

However, as a simple thought experiment shows, such a naive analysis is
inappropriate.  Imagine a local gravitational mass detector $D$ positioned
close to a large mass $M$ which is suspended from a pole. The
decay of an alpha particle will sever the cord  by which the mass
is held to the pole and the mass will drop.  According to the semiclassical
prescription
\be
G_{\mu\nu} = 8 \pi G < \psi | T_{\mu\nu} | \psi > \, , 
\ee
the metric ({\it i.e.}, the mass measured) will slowly decrease.  This is
obviously not what happens.  The mass detector shows a signal which corresponds
to one of the classical trajectories which make up the state $| \psi >$, a
trajectory corresponding to a sudden drop in the gravitational force measured.

The origin of classical density perturbations as a consequence of quantum
fluctuations in a homogeneous state $| \psi >$ can be analyzed along similar
lines.  The quantum to classical transition is picking out$^{\cite{Zurek,qcl1,qcl2}}$ one of
the typical classical trajectories which make up the wave function of $| \psi
>$.  We can implement$^{\cite{Bardeen2,RB84}}$ the procedure as follows: Define a classical scalar field
\be
\varphi_{cl} (\underline{x} , t) = \varphi_0 (t) + \delta \varphi
(\underline{x} , t) 
\ee
with vanishing spatial average of $\delta \varphi$.  The induced classical
energy momentum tensor $T^{cl}_{\mu\nu} (\underline{x}, t)$ which is the source
for the metric is given by
\be \label{Tmunu}
T^{cl}_{\mu\nu} (\underline{x}, t) = T_{\mu\nu} (\varphi_{cl} (\underline{x},
t) ) \, , 
\ee
where $T_{\mu\nu} \, (\varphi_{cl} (\underline{x}, t))$ is defined as the
canonical energy-momentum tensor of the classical scalar field $\varphi_{cl}
(\underline{x}, t)$.  Unless $\delta \varphi$ vanishes, $T^{cl}_{\mu\nu}$ is
inhomogeneous.

For applications to chaotic inflation, we take $| \psi >$ to be a Gaussian
state with mean value $\varphi_0 (t)$
\be
< \psi | \varphi^2 (\underline{x}, t) | \psi > = \varphi_0^2 (t) \, . 
\ee
Its width is taken to be the width of the vacuum state of the free scalar field
theory with mass determined by the curvature of $V(\varphi)$ at $\varphi_0$.
This state is used to define the Fourier transform $\delta \tilde \varphi
(k,t)$ by
\be \label{pertexp}
| \delta \tilde \varphi (k) |^2 = < \psi | \, | \tilde \varphi (k) |^2 \, |
\psi > \, . 
\ee
The amplitude of $\delta \tilde \varphi (k)$ is identified with the width of
the ground state wave function of the harmonic osciallator $\tilde \varphi
(k)$.  (Recall that each Fourier mode of a free scalar field is a harmonic
oscillator).  Note that no divergences arise in the above construction.   

By linearizing (\ref{Tmunu}) about $\varphi_0 (t)$ we obtain the perturbation of the
energy-momentum tensor.  In particular, the energy density fluctuation $\delta
\tilde \rho (k)$ is given by
\be \label{deltarho}
\delta \tilde \rho (k) = \dot \varphi_0 \delta \dot {\tilde \varphi} (k) +
V^\prime (\varphi_0) \delta \tilde \varphi (k) \, . 
\ee
To obtain the initial amplitude (\ref{inpert}) of $\delta M / M$, the above is to be evaluated at the time $t_i (k)$.

The computation of the spectrum of density perturbations produced in the de
Sitter phase has been reduced to the evaluation of the expectation value
(\ref{pertexp}).  First, we must specify the state $| \psi >$.  (Recall that
in non-Minkowski space-times there is no uniquely defined vacuum state of a
quantum field theory$^{\cite{Birrell}}$).  We pick the FRW frame of the pre-inflationary
period.  In this frame, the number density of particles decreases
exponentially.  Hence we choose $| \psi >$ to be the vacuum state in this
frame (see \cite{BH86} for a discussion of other choices).  $\psi [ \tilde \varphi
( \underline{k}), t]$, the wave functional of $|\psi >$, can be calculated
explicitly.  It is basically the superposition of the ground state
wave functions for all oscillators
\be
\psi [ \tilde \varphi ( \underline{k}), t] = N \exp \left\{ - {1\over 2} (2
\pi)^{-3} a^3 (t) \int d^3 k \omega ( \underline{k}, t) | \tilde \varphi
( \underline{k})|^2 \right\} \, . 
\ee
$N$ is a normalization constant and $\omega ( \underline{k}, t) \sim H$ at $t =
t_i (k)$.  Hence
\be \label{deltaphi}
\delta \tilde \varphi ( \underline{k}, t) = (2 \pi)^{3/2}  a^{-3/2}
\omega ( \underline{k}, t)^{-1/2} \sim (2 \pi)^{3/2}  k^{-3/2} H \, , \,
t = t_i (k)\, . 
\ee

Given the above determination of the intitial amplitude of density
perturbations at the time when they leave the Hubble radius during the de
Sitter phase, and the general relativistic analysis of the evolution of
fluctuations discussed in the previous subsection, it is easy to evaluate the r.m.s. inhomogeneities when they reenter the Hubble radius at time $t_f (k)$.

First, we combine (\ref{deltarho}), (\ref{deltaphi}), (\ref{inpert}), (\ref{powersp}) and (\ref{masspow}) to obtain
\be \label{inmass}
\left( {\delta M\over M} \right)^2 \, (k, t_i (k)) \sim k^3 \left({V^\prime
(\varphi_0) \delta \tilde \varphi (k, t_i (k))\over \rho_0} \right)^2 \sim
\left({V^\prime (\varphi_0) H\over \rho_0 } \right)^2 \, . 
\ee
 
If the background scalar field is rolling slowly, then
\be \label{slowroll1}
V^\prime (\varphi_0 (t_i (k))) =  3 H | \dot \varphi_0 (t_i (k)) | \, .
\ee
and
\be \label{slowroll2}
(1 + p/\rho)(t_i(k)) \, \simeq \, \rho_0^{-1} {\dot \varphi_0^2}(t_i(k)) \, .
\ee
Combining (\ref{inmass}), (\ref{slowroll1}), (\ref{slowroll2}) and (\ref{amplpert}) we get
\be
{\delta M\over M} (k, \, t_f (k))  \sim \, {3 H^2 | \dot \varphi_0
(t_i (k)) |\over{\dot \varphi^2_0 (t_i (k))}} =  {3H^2\over{| \dot \varphi_0 (t_i (k))|}} 
\ee
This result can now be evaluated for specific models of inflation to find the
conditions on the particle physics parameters which give a value
\be \label{obs2}
{\delta M\over M} (k, \, t_f (k))  \sim 10^{-5} 
\ee
which is required if quantum fluctuations from inflation are to provide the
seeds for galaxy formation and agree with the CMB anisotropy limits.

For chaotic inflation with a potential
\be
V (\varphi) = {1\over 2} m^2 \varphi^2 \, , 
\ee
we can solve the slow rolling equations for the inflaton to obtain
\be \label{massconstr}
{\delta M\over M} (k, t_f (k))  \sim 10 {m\over m_{pl}} 
\ee
which implies that $m \sim 10^{13} \, {\rm GeV}$ to agree with (\ref{obs2}).

Similarly, for a quartic potential  
\be
V (\varphi) = {1\over 4} \lambda \varphi^4 
\ee
we obtain
\be \label{lambdaconstr}
{\delta M\over M} (k, \, t_f (k)) \sim  10 \cdot \lambda^{1/2} 
\ee
which requires $\lambda \leq 10^{-12}$ in order not to conflict with observations.

The conditions (\ref{massconstr}) and (\ref{lambdaconstr}) require the presence of small parameters in the particle physics model.  
It  has been shown quite generally$^{\cite{Freese}}$ that
small parameters are required if inflation is to solve the fluctuation problem.

I have chosen to present the analysis of fluctuations in inflationary cosmology
in two separate steps in order to highlight the crucial physics issues.
Having done this, it is possible to step back and construct a unified
analysis of the quantum generation and classical evolution of perturbations in an inflationary Universe (for a detailed review see \cite{MFB92}).

The basic point is that at the linearized level, the equations describing both gravitational and matter perturbations can be quantized in a consistent way. The use of gauge invariant variables makes the analysis both physically clear and computationally simple. 

The first step of this analysis is to consider the action for the linear perturbations in a background homogeneous and isotropic Universe, and to express the result in terms of gauge invariant variables describing the fluctuations. Focusing on the scalar perturbations, it turns out that after a lot of algebra the action reduces to the action of a single gauge invariant free scalar field with a time dependent mass$^{\cite{Mukh88,Sasaki}}$ (the time dependence relects the expansion of the background space-time). This result is not suprising. Based on the study of classical cosmological perturbations, we know that there is only one field degree of freedom for the scalar perturbations (for matter theories which obey (\ref{diagonal})). Since at the linearized level there are no mode interactions, the action for this field must be that of a free scalar field. 

The action thus has the same form as the action for a scalar matter field in a time dependent gravitational or electromagnetic background, and we can use standard methods to quantize this theory (see e.g. \cite{Birrell}). If we employ canonical quantization, then the mode functions of the field operator obey the same classical equations as we derived in the gauge-invariant analysis of relativistic perturbations. 

The time dependence of the mass is relected in the nontrivial form of the solutions of the mode equations. The mode equations have growing modes which correspond to particle production or equivalently to the generation and amplification of fluctuations. We can start the system off (e.g. at the beginning of inflation) in the vacuum state (defined as a state with no particles with respect to a local comoving observer). The state defined this way will not be the vacuum state from the point of view of an observer at a later time. The Bogoliubov mode mixing technique can be used to calculate the number density of perticles at a later time. In particular, expectation values of field operators such as the power spectrum can be computed, and the results agree with those obtained using the heuristic analysis at the beginning of this subsection. 
\subsection{Summary}

To summarize the main results of the analysis of density
fluctuations in inflationary cosmology:
\begin{enumerate}
\item{} Quantum vacuum fluctuations in the de Sitter phase of an inflationary
Universe are the source of perturbations.
\item{} The quantum perturbations decohere on scales outside the Hubble
radius and can hence be treated classically.
\item{} As a consequence of the change in the background equation of state, the classical evolution outside the Hubble radius produces a large
amplification of the perturbations.  In fact, unless the particle physics
model contains very small coupling constants, the predicted fluctuations are
in excess of those allowed by the bounds on cosmic microwave anisotropies.
\item{} The quantum generation and classical evolution of fluctuations can be treated in a unified manner. The formalism is no more complicated that the study of a free scalar field in a time dependent background.
\item{} Inflationary Universe models generically produce an approximately scale invariant Harrison-Zel'dovich spectrum
\be
{\delta M\over M} (k , t_f (k) ) \, \simeq \, {\rm const.} 
\ee
\end{enumerate}

It is not hard to construct models which give a different spectrum.  All that
is required is a significant change in $H$ during the period of
inflation. 

Once inside the Hubble radius, the evolution of the mass perturbations is
influenced by the damping effects discussed in Section 3.B, which in turn
depend on the composition of the dark matter.  

On scales which enter the Hubble radius before $t_{eq}$, the perturbations can
only grow logarithmically in time between $t_f(k)$ and $t_{eq}$. This implies
that (up to logarithmic corrections), the mass perturbation spectrum is flat
for wavelengths smaller than $\lambda_{eq}$, the comoving Hubble radius at
$t_{eq}$:
\be \label{Zel1}
{{\delta M} \over M} (\lambda, t) \simeq {\rm const}, \,\,\,\,\,\,\,
t \leq t_{eq},  \, \lambda < \lambda_{eq}, 
\ee
whereas on larger scales
\be \label{Zel2}
{{\delta M} \over M} (\lambda, t) \propto \lambda^{-2}. 
\ee
Equations (\ref{Zel1}) and (\ref{Zel2}) give the power spectrum in an 
$\Omega = 1$ inflationary CDM model.
If the dark matter is hot, then neutrino free streaming cuts off the power
spectrum at $\lambda_J^{max}$ (see (\ref{freestream})).

\section{Topological Defects, Structure Formation and Baryogenesis}

\subsection{Quantifying Data on Large-Scale Structure}

It is length scales corresponding to galaxies and larger which are of greatest
interest in cosmology when attempting to find an imprint of the primordial
fluctuations produced by particle physics.  On these scales, gravitational
effects are assumed to be dominant, and the fluctuations are not too far from
the linear regime.  On smaller scales, nonlinear gravitational and
hydrodynamical effects determine the final state and mask the initial
perturbations.

To set the scales, consider the mean separation of galaxies, which is about
5$h^{-1}$ Mpc$^{\cite{galaxy}}$, and that of Abell clusters which is around 25$h^{-1}$
Mpc$^{\cite{cluster}}$.  The largest coherent structures seen in current redshift surveys have a length of about 
100$h^{-1}$ Mpc$^{\cite{CFA,LCRS}}$, the recent detections of CMB
anisotropies probe the density field on length scales of about $10^3 h^{-1}$
Mpc, and the present horizon corresponds to a distance of about $3 \cdot 10^3
h^{-1}$ Mpc.

Galaxies are gravitationally bound systems containing billions of stars.  They
are non-randomly distributed in space.  A quantitative measure of this
non-randomness is the ``two-point correlation function" $\xi_2 (r)$ which gives
the excess probability of finding a galaxy at a distance $r$ from a given
galaxy:
\be
\xi_2 (r) = < \, {n (r) - n_0\over n_0} \, >  \, . 
\ee
Here, $n_0$ is the average number density of galaxies, and $n(r)$ is the
density of galaxies a distance $r$ from a given one.  The pointed braces stand
for ensemble averaging.

Recent observational results from a various galaxy redshift surveys yield reasonable agreement$^{\cite{twopoint}}$ with a form 
\be
\xi_2 (r) \simeq \left({r_0\over r} \right)^\gamma 
\ee
with scaling length $r_0 \simeq 5 h^{-1}$ Mpc and power $\gamma \simeq 1.8$.  A
theory of structure formation must explain both the amplitude and the slope of
this correlation function. 

On scales larger than galaxies, a better way to quantify structure is by means of large-scale systematic redshift surveys. Such surveys have
discovered coherent  planar structures and voids on scales of up
to $100 h^{-1}$ Mpc. Fig. 12 is a
sketch of redshift versus angle $\alpha$ in the sky for one $6^o$ slice of the sky$^{\cite{CFA}}$.  The second
direction in the sky has been projected onto the $\alpha -z$ plane.  The most
prominent feature is the band of galaxies at a distance of about $100h^{-1}$
Mpc.  This band also appears in neighboring slices and is therefore presumably
part of a planar density enhancement of comoving planar size of at least $(50
\times 100) \times h^{-2}$ Mpc$^2$.  This structure is often called the ``Great
Wall."  The challenge for theories of structure formation is not only to explain the fact that galaxies are nonrandomly distributed, but also to predict both
the observed scale and topology of the galaxy distribution. Topological defect models of structure formation attempt to address these questions.

Until 1992 there was little evidence for any convergence of the galaxy
distribution towards homogeneity.   Each new survey led to the discovery of new
coherent structures in the Universe on a scale comparable to that of the
survey.  In 1996, results of a much deeper redshift survey were
published$^{\cite{LCRS}}$ which for the first time find no coherent structures on the scale of the entire survey. In fact, so coherent structures on 
scales larger than $100 h^{-1}$ Mpc are seen.  This is the first direct evidence for the cosmological principle from optical surveys (the isotropy of the CMB has for a long time been a strong point in its support).

\begin{figure}
\begin{center}
\leavevmode
\epsfxsize=10cm \epsfbox{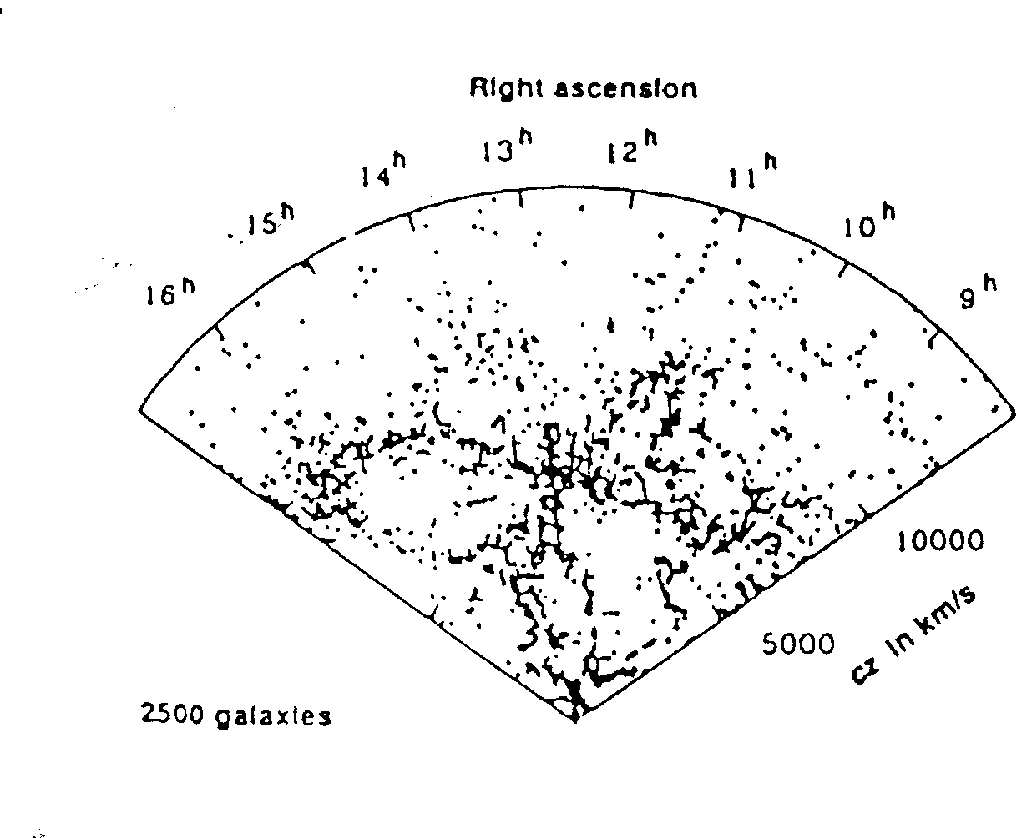}
\caption{ 
Results from the CFA redshift survey. Radial distance
gives the redshift of galaxies, the angular distance corresponds to right
ascension. The results from several slices of the sky (at different
declinations) have been projected into the same cone.}
\end{center}
\end{figure}

In summary, a lot of data from optical and infrared galaxies are
currently available, and new data are being collected at a rapid rate.  The
observational constraints on theories of structure formation are becoming
tighter.   

\subsection{Toplogical Defects}

According to particle physics theories, matter at high energies and
temperatures must be described in terms of fields.  Gauge symmetries have
proved to be extremely useful in describing the standard model of particle
physics, according to which at high energies the laws of nature are invariant
under a nonabelian group  $G$ of internal symmetry transformations
$G = {\rm SU} (3)_c \times {\rm SU} (2)_L \times U(1)_Y$
which at a temperature of about 200 MeV is spontaneously broken down to
$G^\prime = {\rm SU} (3)_c \times {\rm U} (1)$. The subscript on the SU(3) subgroup indicates that it is the color symmetry
group of the strong interactions, ${\rm SU} (2)_L \times $ U(1)$_Y$ is the
Glashow-Weinberg-Salam (WS) model of weak and electromagnetic interactions, the
subscripts $L$ and $Y$ denoting left handedness and hypercharge respectively.
At low energies, the WS model spontaneously breaks to the U(1) subgroup of
electromagnetism.

Spontaneous symmetry breaking is induced by an order parameter $\varphi$ taking
on a nontrivial expectation value $< \varphi >$ below a certain temperture
$T_c$.  In some particle physics models, $\varphi$ is a fundamental scalar
field in a nontrivial representation of the gauge group $G$ which is broken.
However, $\varphi$ could also be a fermion condensate, as in the BCS theory of superconductivity.

Earlier we have seen that symmetry breaking phase
transitions in gauge field theories do not, in general, lead to
inflation.  In most models, the coupling constants which arise in the
effective potential for the scalar field $\varphi$ driving the phase
transition are too large to generate a period of slow rolling which
lasts more than one Hubble time $H^{-1} (t)$.  Nevertheless, there are
interesting remnants of the phase transition: topological defects.
 
Consider a single component real scalar field with a typical symmetry breaking
potential
\be \label{stringpot}
V (\varphi) = {1\over 4} \lambda (\varphi^2 - \eta^2)^2 
\ee
Unless $\lambda \ll 1$ there
will be no inflation.  The phase transition will take place on a short time
scale $\tau < H^{-1}$, and will lead to correlation regions of radius $\xi <
t$ inside of which $\varphi$ is approximately constant, but outside of which
$\varphi$ ranges randomly over the vacuum manifold ${\cal M}$, the set of
values
of $\varphi$ which minimizes $V(\varphi)$ -- in our example $\varphi
= \pm \eta$.  The correlation regions are separated by domain walls, regions in
space where $\varphi$ leaves the vacuum manifold ${\cal M}$ and where,
therefore, potential energy is localized.  Via the usual gravitational
force, this energy density can act as a seed for structure.

Topological defects are familiar from solid state and condensed matter
systems.  Crystal defects, for example, form when water freezes or
when a metal crystallizes$^{\cite{Mermin}}$.  Point defects, line defects and planar
defects are possible.  Defects are also common in liquid crystals$^{\cite{LQ}}$.
They arise in a temperature quench from the disordered to the ordered
phase.  Vortices in $^4$He are analogs of global cosmic strings.
Vortices and other defects are also produced$^{\cite{Salomaa}}$ during a quench below the
critical temperature in $^3$He.  Finally, vortex lines may play an
important role in the theory of superconductivity$^{\cite{Abrikosov}}$.

The analogies between defects in particle physics and condensed matter
physics are quite deep.  Defects form for the same reason: the vacuum
manifold is topologically nontrivial.  The arguments$^{\cite{Kibble1}}$ which say that in
a theory which admits defects, such defects will inevitably form, are
applicable both in cosmology and in condensed matter physics.
Different, however, is the defect dynamics.  The motion of defects in
condensed matter systems is friction-dominated, whereas the defects in
cosmology obey relativistic equations, second order in time
derivatives, since they come from a relativistic field theory.

After these general comments we turn to a classification of
topological defects$^{\cite{Kibble1}}$.  We consider theories with an $n$-component order
parameter $\varphi$ and with a potential energy function (free energy
density) of the form (6.1) with $\varphi^2 = \sum\limits^n_{i = 1} \, \varphi^2_i$. 

There are various types of local and global topological defects
(regions of trapped energy density) depending on the number $n$ of components
of $\varphi$ (see e.g. \cite{VilShell} for a comprehensive survey of topological defect models).
The more rigorous mathematical definition refers to the homotopy
of ${\cal M}$.  The words ``local" and ``global" refer to whether the symmetry
which is broken is a gauge or global symmetry.  In the case of local
symmetries, the topological defects have a well defined core outside of which
$\varphi$ contains no energy density in spite of nonvanishing gradients
$\nabla \varphi$:  the gauge fields $A_\mu$ can absorb the gradient,
{\it i.e.,} $D_\mu \varphi = 0$ when $\partial_\mu \varphi \neq 0$,
where the covariant derivative $D_\mu$ is defined by
$D_\mu = \partial_\mu + ie \, A_\mu$, $e$ being the gauge coupling constant.
Global topological defects, however, have long range density fields and
forces.
\par
Table 1 contains a list of topological defects with their topological
characteristics.  A ``v" marks acceptable theories, a ``x" theories which are
in conflict with observations (for $\eta \sim 10^{16}$ GeV).

\begin{figure}
\begin{center}
\leavevmode
\epsfxsize=12cm \epsfbox{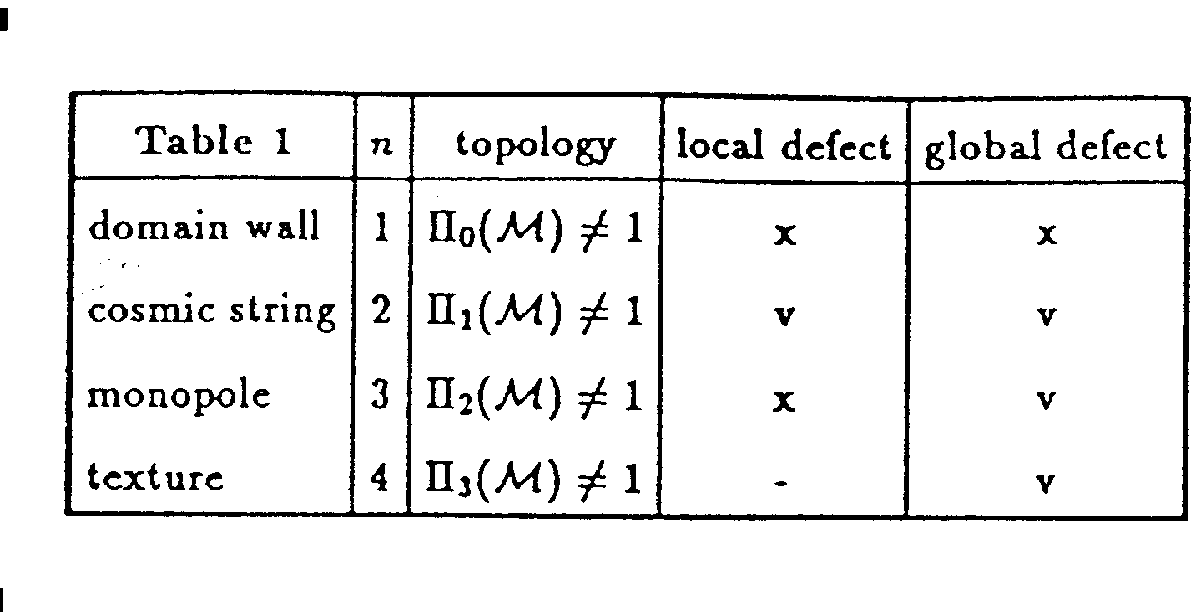}
\end{center}
\end{figure}
  
Theories with domain walls are ruled out$^{\cite{nodw}}$ since a single domain wall stretching
across the Universe today would overclose the Universe.  Local monopoles are
also ruled out$^{\cite{nomon}}$ since they would overclose the Universe.  Local
textures are ineffective at producing structures because there is no traped potential energy.

From now on we will focus on one type of defects: cosmic strings (see e.g. \cite{VilShell,HK95,RB94} for recent reviews, and \cite{Vil85} for a classic review paper). These arise
in theories with a complex order parameter $(n = 2)$. In this case the vacuum manifold of the model is
\be
{\cal M} = S^1 \, , 
\ee
which has nonvanishing first homotopy group:
\be
\Pi_1 ({\cal M}) = Z \neq 1 \, . 
\ee
A cosmic string is a line of trapped energy density which arises
whenever the field $\varphi (x)$ circles ${\cal M}$ along a closed path
in space ({\it e.g.}, along a circle).  In this case, continuity of
$\varphi$ implies that there must be a point with $\varphi = 0$ on any
disk whose boundary is the closed path.  The points on different sheets
connect up to form a line overdensity of field energy.  

To construct a field configuration with a string along the $z$ axis$^{\cite{Nielsen}}$,
take $\varphi (x)$ to cover ${\cal M}$ along a circle with radius $r$
about the point $(x,y) = (0,0)$:
\be \label{stringconf}
\varphi (r, \vartheta ) \simeq \eta e^{i \vartheta} \, , \, r \gg
\eta^{-1} \, . 
\ee
This configuration has winding number 1, {\it i.e.}, it covers ${\cal
M}$ exactly once.  Maintaining cylindrical symmetry, we can extend
(\ref{stringconf}) to arbitrary $r$
\be
\varphi (r, \, \vartheta) = f (r) e^{i \vartheta} \, , 
\ee
where $f (0) = 0$ and $f (r)$ tends to $\eta$ for large $r$.  The
width $w$ can be found by balancing potential and tension
energy.  The result is 
\be
w \, \sim \, \lambda^{-1/2} \eta^{-1} \, .
\ee

For local cosmic strings, {\it i.e.}, strings arising due to the
spontaneous breaking of a gauge symmetry, the energy density decays
exponentially for $r \gg w$.  In this case, the energy $\mu$
per unit length of a string is finite and depends only on the symmetry
breaking scale $\eta$
\be
\mu \sim \eta^2 
\ee
(independent of the coupling $\lambda$).  The value of $\mu$ is the
only free parameter in a cosmic string model.

\subsection{Formation of Defects in Cosmological Phase Transitions}

The symmetry breaking phase transition takes place at $T = T_c$ (called the critical temperature).  From condensed matter physics it is well
known that in many cases topological defects form during phase transitions,
particularly if the transition rate is fast on a scale compared to the system
size.  When cooling a metal, defects in the crystal configuration will be
frozen in; during a temperature quench of $^4$He, thin vortex tubes of the
normal phase are trapped in the superfluid; and analogously in a temperature
quench of a superconductor, flux lines are trapped in a surrounding sea of the
superconducting Meissner phase.

In cosmology, the rate at which the phase transition proceeds is given by the
expansion rate of the Universe.  Hence, topological defects will inevitably be
produced in a cosmological phase transition$^{\cite{Kibble1}}$, provided the underlying particle physics model allows such defects. 
 
The argument which  ensures that in theories which admit
topological or semitopological defects, such defects will be produced
during a phase transition in the very early Universe is called the Kibble mechanism$^{\cite{Kibble1}}$. To illustrate the physics, consider
a mechanical toy model, first introduced by Mazenko, Unruh
and Wald$^{\cite{MUW}}$. Take (see
Fig. 13) a lattice of points on a flat table.  At each point, a pencil
is pivoted.  It is free to rotate and oscillate.  The tips of nearest
neighbor pencils are connected with springs (to mimic the spatial
gradient terms in the scalar field Lagrangean).  Newtonian gravity
creates a potential energy $V(\varphi)$ for each pencil ($\varphi$ is
the angle relative to the vertical direction).  $V(\varphi)$ is
minimized for $| \varphi | = \eta$ (in our toy model $\eta = \pi /
2$).  Hence, the Lagrangean of this pencil model is analogous to that
of a scalar field with symmetry breaking potential (\ref{stringpot}).

\begin{figure}
\begin{center}
\leavevmode
\epsfxsize=10cm \epsfbox{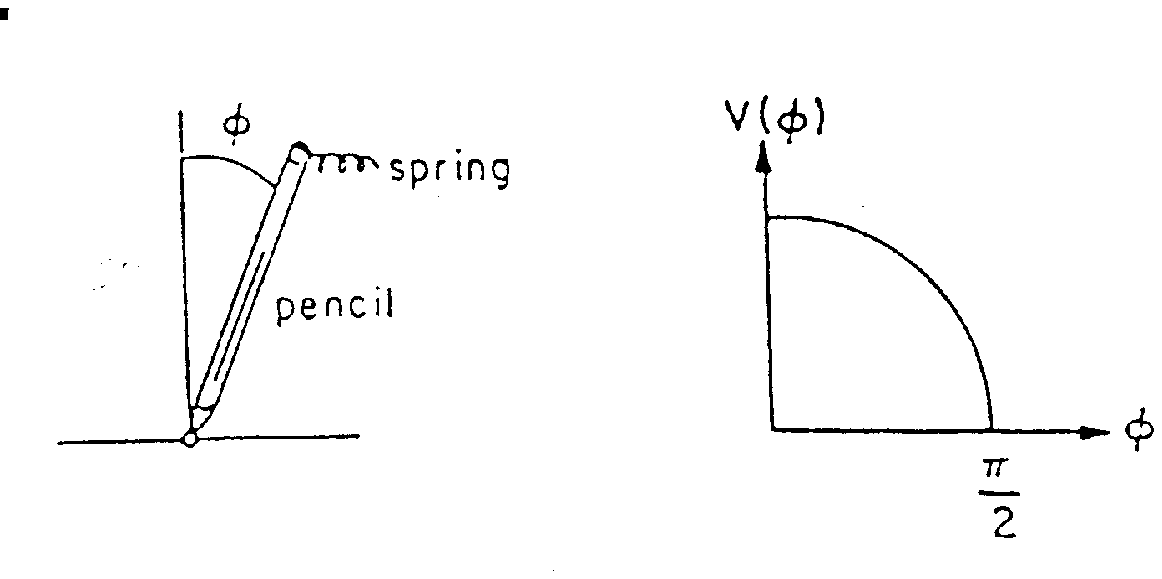}
\caption{
The pencil model: the potential energy of a
simple pencil has the same form as that of scalar fields used for
spontaneous symmetry breaking.  The springs connecting nearest
neighbor pencils give rise to contributions to the energy which mimic
spatial gradient terms in field theory.}
\end{center}
\end{figure}

At high temperatures $T \gg T_c$, all pencils undergo large amplitude
high frequency oscillations.  However, by causality, the phases of
oscillation of pencils with large separation $s$ are uncorrelated.
For a system in thermal equilibrium, the length $s$ beyond which
phases are random is the correlation length $\xi (t)$.  However, since
the system is quenched rapidly, there is a causality bound on
$\xi$:
\be
\xi (t) < t \, , 
\ee
where $t$ is the causal horizon.

The critical temperature $T_c$ is the temperature at which the
thermal energy is equal to the energy a pencil needs to jump from
horizontal to vertical position.  For $T < T_c$, all pencils want to
lie flat on the table.  However, their orientations are random beyond
a distance $\xi (t)$ determined by equating the free energy gained by
symmetry breaking (a volume effect) with the gradient energy lost (a surface
effect).  As expected, $\xi (T)$ diverges at $T_c$. Very close to $T_c$, the
thermal energy $T$ is larger than the volume energy gain $E_{corr}$ in a
correlation volume. Hence, these domains are unstable to thermal fluctuations.
As $T$ decreases, the thermal energy decreases more rapidly than $E_{corr}$.
Below the ``Ginsburg temperature" $T_G$, there
is insufficient thermal energy to excite a correlation volume into the
state $\varphi = 0$.  Domains of size
\be \label{corrlength}
\xi (t_G) \sim \lambda^{-1} \eta^{-1} 
\ee
freeze out$^{\cite{Kibble1,Kibble2}}$.  The boundaries between these domains become
topological defects. An improved version of this argument has recently been given by Zurek$^{\cite{Zurek2}}$ (see also \cite{BDH}).

We conclude that in a theory in which a symmetry breaking phase
transitions satisfies the topological criteria for the existence of a
fixed type of defect, a network of such defects will form during the
phase transition and will freeze out at the Ginsburg temperature.  The
correlation length is initially given by (\ref{corrlength}), if the field
$\varphi$ is in thermal equilibrium before the transition.
Independent of this last assumption, the causality bound implies that
$\xi (t_G) < t_G$.

For times $t > t_G$ the evolution of the network of defects may be
complicated (as for cosmic strings) or trivial (as for textures).  In
any case (see the caveats of \cite{caveat}), the causality bound
persists at late times and states that even at late times, the mean
separation and length scale of defects is bounded by $\xi (t) \leq t$.

Applied to cosmic strings, the Kibble mechanism implies that at the
time of the phase transition, a network of cosmic strings with typical
step length $\xi (t_G)$ will form.  According to numerical
simulations$^{\cite{VachVil}}$, about 80\% of the initial energy is in infinite
strings (strings with curvature radius larger than the Hubble radius) and 20\% in closed loops.

\subsection{Evolution of Strings and Scaling}

\begin{figure}
\begin{center}
\leavevmode
\epsfxsize=10cm \epsfbox{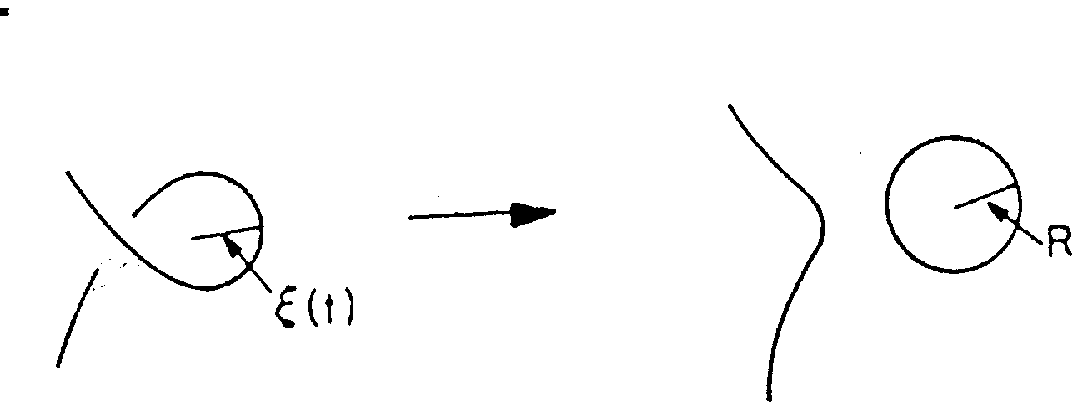}
\caption{
Formation of a loop by a self intersection of an
infinite string. According to the original cosmic string scenario, loops form
with a radius $R$ determined by the instantaneous coherence length of the
infinite string network.}
\end{center}
\end{figure}

The evolution of the cosmic string network for $t > t_G$ is
complicated.  The key processes are loop production
by intersections of infinite strings (see Fig. 14) and loop shrinking
by gravitational radiation.  These two processes combine to create a
mechanism by which the infinite string network loses energy (and
length as measured in comoving coordinates).  It can be shown (see e.g. \cite{Vil85}) that, as
a consequence, the correlation length of the string network is always
proportional to its causality limit
\be
\xi (t) \sim t \, . 
\ee
Hence, the energy density $\rho_\infty (t)$ in long strings is a fixed
fraction of the background energy density $\rho_c (t)$
\be
\rho_\infty (t) \sim \mu \xi (t)^{-2} \sim \mu t^{-2} 
\ee
or
\be
{\rho_\infty (t)\over{\rho_c (t)}} \sim G \mu \, . 
\ee

We conclude that the cosmic string network approaches a ``scaling
solution" in which the statistical properties of the
network are time independent if all distances are scaled to the
horizon distance.
  
\subsection {Cosmic Strings and Structure Formation}

The starting point of the structure formation scenario in the cosmic
string theory is the scaling solution for the cosmic string network,
according to which at all times $t$ (in particular at $t_{eq}$, the
time when perturbations can start to grow) there will be a few long
strings crossing each Hubble volume, plus a distribution of loops of
radius $R \ll t$ (see Fig. 15).

The cosmic string model admits three mechanisms for structure
formation:  loops, filaments, and wakes.  Cosmic string loops have the same
time averaged field as a point source with mass$^{\cite{Turok84}}$
$ M (R) = \beta R \mu $, $R$ being the loop radius and $\beta \sim 2 \pi$.  Hence, loops will be seeds for spherical accretion of dust and radiation.

For loops with $R \leq t_{eq}$, growth of perturbations in a model
dominated by cold dark matter starts at $t_{eq}$.  Hence, the mass at
the present time will be
\be
M (R, \, t_0) = z (t_{eq}) \beta \, R \mu \, . 
\ee

\begin{figure}
\begin{center}
\leavevmode
\epsfxsize=8cm \epsfbox{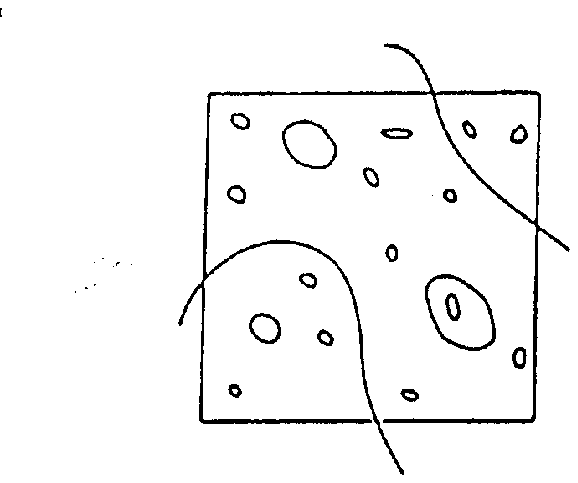}
\caption{ 
Sketch of the scaling solution for the
cosmic string network.  The box corresponds to one Hubble volume at
arbitrary time $t$.}
\end{center}
\end{figure}

In the original cosmic string model$^{\cite{ZelVil,TB86}}$ it was assumed
that loops dominate over wakes.  However, according to the newer cosmic string evolution simulations$^{\cite{CSsimuls}}$, most of the energy in strings is in the long strings, and hence the loop accretion mechanism is subdominant. 

The second mechanism involves long strings moving with relativistic
speed in their normal plane which give rise to
velocity perturbations in their wake$^{\cite{SilkVil}}$.  The mechanism is illustrated in
Fig. 16:
space normal to the string is a cone with deficit angle$^{\cite{Vil81}}$
\be \label{deficit}
\alpha = 8 \pi G \mu \, . 
\ee
If the string is moving with normal velocity $v$ through a bath of dark
matter, a velocity perturbation
\be
\delta v = 4 \pi G \mu v \gamma  
\ee
[with $\gamma = (1 - v^2)^{-1/2}$] towards the plane behind the string
results.  At times after $t_{eq}$, this induces planar overdensities,
the most
prominent ({\it i.e.}, thickest at the present time) and numerous of which were
created at $t_{eq}$, the time of equal matter and
radiation$^{\cite{TV86,AS87,LP90}}$.  The
corresponding planar dimensions are (in comoving coordinates)
\be
t_{eq} z (t_{eq}) \times t_{eq} z (t_{eq}) v \sim (40 \times 40 v) \,
{\rm Mpc}^2\, . 
\ee

\begin{figure}
\begin{center}
\leavevmode
\epsfxsize=14cm \epsfbox{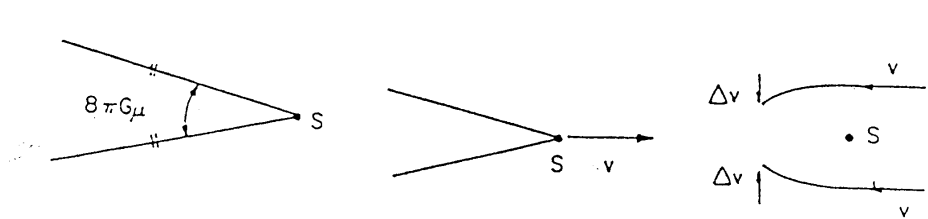}
\caption{
Sketch of the mechanism by which a long
straight cosmic string $S$ moving with velocity $v$ in transverse
direction through a plasma induces a velocity perturbation $\Delta v$
towards the wake. Shown on the left is the deficit angle, in the
center is a sketch of the string moving in the plasma, and on the
right is the sketch of how the plasma moves in the frame in which the
string is at rest.}
\end{center}
\end{figure}
  
The thickness $d$ of these wakes can be calculated using the
Zel'dovich approximation$^{\cite{Zeld70}}$.  The result is (for $G \mu = 10^{-6}$)
\be
d \simeq G \mu v \gamma (v) z (t_{eq})^2 \, t_{eq} \simeq 4 v \, {\rm
Mpc} \, . 
\ee
 
Wakes arise if there is little small scale structure on the string.
In this case, the string tension equals the mass density, the string
moves at relativistic speeds, and there is no local gravitational
attraction towards the string.

In contrast, if there is small scale structure on strings,
then the string tension $T$ is smaller$^{\cite{Carter}}$ than the mass per unit
length $\mu$ , and thus there
is a gravitational force towards the string which gives rise to
cylindrical accretion, producing filaments$^{\cite{fils}}$.

Which of the mechanisms -- filaments or wakes -- dominates is
determined by the competition between the velocity induced by the Newtonian gravitational potential of the strings and the velocity perturbation of the wake.   

The cosmic string model predicts a scale-invariant spectrum of density
perturbations, exactly like inflationary Universe models but for a
rather different reason.  Consider the {\it r.m.s.} mass fluctuations
on a length scale $2 \pi k^{-1}$ at the time $t_H (k)$ when this scale
enters the Hubble radius.  From the cosmic string scaling solution it
follows that a fixed ({\it i.e.}, $t_H (k)$ independent) number
$\tilde v$ of strings of length of the order $t_H (k)$ contribute to
the mass excess $\delta M (k, \, t_H (k))$.  Thus
\be
{\delta M\over M} \, (k, \, t_H (k)) \sim \, {\tilde v \mu t_H
(k)\over{G^{-1} t^{-2}_H (k) t^3_H (k)}} \sim \tilde v \, G \mu \, .
\ee
Note that the above argument predicting a scale invariant spectrum
will hold for all topological defect models which have a scaling
solution, in particular also for global monopoles and textures.

The amplitude of the {\it r.m.s.} mass fluctuations (equivalently: of
the power spectrum) can be used to normalize $G \mu$.  Since today on
galaxy cluster scales
\be
{\delta M\over M} (k, \, t_0) \sim 1 \, , 
\ee
the growth rate of fluctuations linear in $a(t)$ yields
\be
{\delta M\over M} \, (k, \, t_{eq}) \sim 10^{-4} \, , 
\ee
and therefore, using $\tilde v \sim 10$,
\be
G \mu \sim 10^{-5} \, . 
\ee
A similar value is obtained by normalizing the model to the COBE amplitude of CMB anisotropies on large angular scales$^{\cite{BBS,Periv}}$ (the normalizations from COBE and from the power spectrum of density perturbations on large scales agree to within a factor of 2).
Thus, if cosmic strings are to be relevant for structure formation,
they must arise due to a symmetry breaking at energy scale $\eta
\simeq 10^{16}$GeV.  This scale happens to be the scale of unification (GUT)
of weak, strong and electromagnetic interactions.  It is tantalizing
to speculate that cosmology is telling us that there indeed was new
physics at the GUT scale.

A big advantage of the cosmic string model over inflationary Universe
models is that HDM is a viable dark matter candidate.  Cosmic string
loops survive free streaming, as discussed in Section 3.B, and can
generate nonlinear structures on galactic scales, as discussed in
detail in \cite{BKST,Edbert}.  Accretion of hot dark matter by a string wake
was studied in \cite{LP90}. In this case, nonlinear perturbations
develop only late.  At some time $t_{nl}$, all scales up to a distance
$q_{\rm max}$ from the wake center go nonlinear.  Here
\be
q_{\rm max} \sim G \mu v \gamma (v) z (t_{eq})^2 t_{eq} \sim 4 v \,
{\rm Mpc} \, , 
\ee
and it is the comoving thickness of the wake at $t_{nl}$.  Demanding
that $t_{nl}$ corresponds to a redshift greater than 1 leads to the
constraint
\be
G \mu > 5 \cdot 10^{-7} \, . 
\ee
Note that in a cosmic string and hot dark matter model, wakes form nonlinear structures only very recently. Accretion onto loops and small scale structure on the long strings provide two mechanisms which may lead to high redshift objects such as quasars and high redshift galaxies. The first mechanism has recently been studied in \cite{MB96}, the second in \cite{AB95,ZLB96}.

The power spectrum of density fluctuations in a cosmic string model with HDM has recently been studied numerically by M\"ah\"onen$^{\cite{Mahonen}}$, based on previous work of \cite{Hara} (see also \cite{AS92} for an earlier semi-analytical study). The spectral shape agrees quite well with observations, and a bias factor of less than 2 is required to give the best-fit amplitude for a COBE normalized model. Note, however, that the results depend quite sensitively on the details of the string scaling solution which are at present not well understood.

Due to lack of space, I will not discuss the global monopole$^{\cite{BenRhie}}$ and global texture$^{\cite{Turok89}}$ models of structure formation. There has been a lot of work on the texture model, and the reader is referred to \cite{Turok91,Durrer94} for recent review articles.

\subsection{Signatures for Strings}

The cosmic string theory of structure formation makes several distinctive predictions, both in terms of the galaxy distribution and in terms of CMB anisotropies. On large scales (corresponding to the comoving Hubble radius at $t_{eq}$ and larger, structure is predicted to be dominated either by planar$^{\cite{TV86,AS87,LP90}}$ or filamentary$^{\cite{fils}}$ galaxy concentrations. For models in which the strings have no local gravity, the resulting nonlinear structures will look very different from the nonlinear structures in models in which local gravity is the dominant force. As discovered and discussed recently in \cite{SB96}, a baryon number excess is predicted in the nonlinear wakes. This may explain the ``cluster baryon crisis"$^{\cite{clusterbaryon}}$, the fact that the ratio of baryons to dark matter in rich clusters is larger than what is compatible with the nucleosynthesis constraints in a spatially flat Universe.

\begin{figure}
\begin{center}
\leavevmode
\epsfxsize=7cm \epsfbox{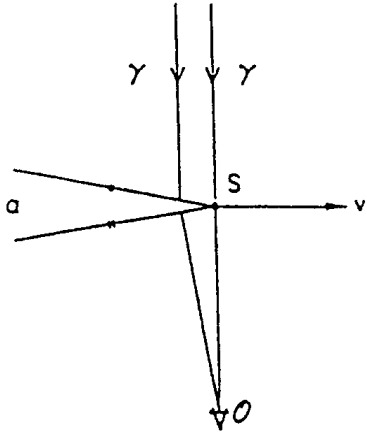}
\caption{
Sketch of the Kaiser-Stebbins effect by
which cosmic strings produce linear discontinuities in the CMB. Photons
$\gamma$ passing on different sides of a moving string $S$ (velocity $v$)
towards the observer ${\cal O}$ receive a relative Doppler shift due to the
conical nature of space perpendicular to the string (deficit angle $\alpha$).}
\end{center}
\end{figure} 

As described in the previous subsection, space perpendicular to a long straight
cosmic string is conical with deficit angle given by (\ref{deficit}).  Consider
now CMB radiation approaching an observer in a direction normal to the
plane spanned by the string and its velocity vector (see Fig. 17).
Photons arriving at the observer having passed on different sides of
the string will obtain a relative Doppler shift which translates into
a temperature discontinuity of amplitude$^{\cite{KS84}}$
\be
{\delta T\over T} = 4 \pi G \mu v \gamma (v) \, , 
\ee
where $v$ is the velocity of the string.  Thus, the distinctive
signature for cosmic strings in the microwave sky are line
discontinuities in $T$ of the above magnitude.

Given ideal maps of the CMB sky it would be easy to detect strings.
However, real experiments have finite beam width.  Taking into account
averaging over a scale corresponding to the beam width will smear out
the discontinuities, and it turns out to be surprisingly hard to
distinguish the predictions of the cosmic string model from that of
inflation-based theories using quantitative statistics which are easy
to evaluate analytically, such as the kurtosis of the spatial gradient
map of the CMB$^{\cite{MPB94}}$. There may be ways to distinguish between string and inflationary models by looking at the angular power spectrum of CMB anisotropies. Work on this subject, however, is still controversial$^{\cite{AAcrew,HuWhite,Turok96}}$.

Global textures also produce distinctive non-Gaussian signatures$^{\cite{TurSper}}$ in CMB maps. In fact, these signatures are more pronounced and on larger scales than the signatures in the cosmic string model. 

\subsection{Principles of Baryogenesis}

Baryogenesis is another area where particle physics and cosmology connect in a very deep way. The goal is to explain the observed asymmetry between matter and antimatter in the Universe. In particular, the objective is to be able to explain the observed value of the net baryon to entropy ratio at the present time
\be
{{\Delta n_B} \over s}(t_0) \, \sim \, 10^{-10} 
\ee
starting from initial conditions in the very early Universe when this ratio vanishes. Here, $\Delta n_B$ is the net baryon number density and $s$ the entropy density.

As pointed out by Sakharov$^{\cite{Sakharov}}$, three basic criteria must be satisfied in order to have a chance at explaining the data:
\begin{enumerate}
\item{} The theory describing the microphysics must contain baryon number violating processes.
\item{} These processes must be C and CP violating.
\item{} The baryon number violating processes must occur out of thermal equilibrium.
\end{enumerate}

As was discovered in the 1970's$^{\cite{GUTBG}}$, all three criteria can be satisfied in GUT theories. In these models, baryon number violating processes are mediated by superheavy Higgs and gauge particles. The baryon number violation is visible in the Lagrangian, and occurs in perturbation theory (and is therefore in principle easy to calculate). In addition to standard model CP violation, there are typically many new sources of CP violation in the GUT sector. The third Sakharov condition can also be realized: After the GUT symmetry-breaking phase transition, the superheavy particles may fall out of thermal equilibrium. The out-of-equilibrium decay of these particles can thus generate a nonvanishing baryon to entropy ratio. 

The magnitude of the predicted $n_B / s$ depends on the asymmetry $\varepsilon$ per decay, on the coupling constant $\lambda$ of the $n_B$ violating processes, and on the ratio $n_X / s$ of the number density $n_X$ of superheavy Higgs and gauge particles to the number density of photons, evaluated at the time $t_d$ when the baryon number violating processes fall out of thermal equilibrium, and assuming
that this time occurs after the phase transition. The quantity $\varepsilon$ is proportional to the CP-violation parameter in the model. In a GUT theory, this CP violation parameter can be large (order 1), whereas in the standard electroweak theory it is given by the CP violating phases in the CKM mass matrix and is very small. As shown in \cite{GUTBG} it is easily possible to construct models which give the right $n_B / s$ ratio after the GUT phase transition (for recent reviews of baryogenesis see \cite{Dolgov} and \cite{RubShap}).
 
\subsection{GUT Baryogenesis and Topological Defects}

The ratio $n_B / s$, however, does not only depend on $\varepsilon$, but also on $n_X / s (t_d)$. If the temperature $T_d$ at the time $t_d$ is greater than the mass $m_X$ of the superheavy particles, then it follows from the thermal history in standard cosmology that $n_X \sim s$. However, if $T_d < m_X$, then the number density of $X$ particles is diluted exponentially in the time interval between when $T = m_X$ and when $T = T_d$. Thus, the predicted baryon to entropy ratio is exponentially suppressed:
\be \label{expdecay}
{n_B \over s} \, \sim \, {1 \over {g^*}} \lambda^2 \varepsilon e^{- m_X / T_d} \, ,
\ee
where $g^*$ is the number of spin degrees of freedom in thermal equilibrium at the time of the phase transition.
In this case, the standard GUT baryogenesis mechanism is ineffective.

However, topological defects may come to the rescue$^{\cite{BDH92}}$. As we have seen earlier in this section, topological defects will inevitably be produced in the symmetry breaking GUT transition provided they are topologically allowed in that symmetry breaking scheme. The topological defects provide an alternative mechanism of GUT baryogenesis.

Inside of topological defects, the GUT symmetry is restored. In fact, the defects can be viewed as solitonic configurations of $X$ particles. The continuous decay of defects at times after $t_d$ provides an alternative way to generate a nonvanishing baryon to entropy ratio. The defects constitute out of equilibrium configurations, and hence their decay can produce a nonvanishing $n_B / s$ in the same way as the decay of free $X$ quanta. 

The way to compute the resulting $n_B / s$ ratio is as follows: The defect
scaling solution gives the energy density in defects at all times. Taking the time derivative of this density, and taking into account the expansion of the Universe, we obtain the loss of energy attributed to defect decay. By energetics, we can estimate the number of decays of individual quanta which the defect decay corresponds to. We can then use the usual perturbative results to compute the resulting net baryon number.

Provided that $m_X < T_d$, then at the time when the baryon number violating processes fall out of equilibrium (when we start generating a nonvanishing $n_B$) the energy density in free $X$ quanta is much larger than the defect density, and hence the defect-driven baryogenesis mechanism is subdominant. However, if $m_X > T_d$, then as indicated in (\ref{expdecay}), the energy density in free quanta decays exponentially. In constrast, the density in defects only
decreases as a power of time, and hence soon dominates baryogenesis.

One of the most important ingredients in the calculation is the time dependence of $\xi(t)$, the separation between defects. Immediately after the phase transition at the time $t_f$ of the formation of the defect network, the separation is $\xi(t_f) \sim \lambda^{-1} \eta^{-1}$. In the time period immediately following, the time period of relevance for baryogenesis, $\xi(t)$ approaches the Hubble radius according to the equation$^{\cite{Kibble2}}$
\be \label{defsep}
\xi(t) \, \simeq \, \xi(t_f) ({t \over {t_f}})^{5/4} \, .
\ee
Using this result to calculate the defect density, we obtain after some algebra
\be \label{barres}
{{n_B} \over s}|_{\rm defect} \, \sim \, \lambda^2 {{T_d} \over \eta} {{n_B} \over s}|_0 \, ,
\ee
where $n_B / s|_0$ is the unsuppressed value of $n_B / s$ which can be obtained using the standard GUT baryogenesis mechanism. We see from (\ref{barres}) that even for low values of $T_d$, the magnitude of $n_B / s$ which is obtained via the defect mechanism is only suppressed by a power of $T_d$. However, the maximum strength of the defect channel is smaller than the maximum strength of the usual mechanism by a geometrical suppression factor $\lambda^2$ which expresses the fact that even at the time of defect formation, the defect network only occupies a small volume.

\subsection{Electroweak Baryogenesis and Topological Defects}

It has been known for some time that there are baryon number violating processes even in the standard electroweak theory. These processes are, however, nonperturbative. They are connected with the t'Hooft anomaly$^{\cite{tHooft}}$, which in turn is due to the fact that the gauge theory vacuum is degenerate, and that the different degenerate vacuum states have different quantum numbers (Chern-Simons numbers). In theories with fermions, this implies different baryon number. Configurations such as sphalerons$^{\cite{sphal}}$ which interpolate between two such vacuum states thus correspond to baryon number violating processes.

As pointed out in \cite{KRS85}, the anomalous baryon number violating processes are in thermal equilibrium above the electroweak symmetry breaking scale. Therefore, any net baryon to entropy ratio generated at a higher scale will be erased, unless this ratio is protected by an additional quantum number such as a nonvanishing $B - L$ which is conserved by electroweak processes.

However, as first suggested in \cite{Shap} and discussed in detail in many recent papers (see \cite{EWBGrev} for reviews of the literature), it is possible to regenerate a nonvanishing $n_B / s$ below the electroweak symmetry breaking scale. Since there are $n_B$ violating processes and both C and CP violation in the standard model, Sakharov's conditions are satisfied provided that one can realize an out-of-equilibrium state after the phase transition. Standard model CP violation is extremely weak. Thus, it appears necessary to add some sector with extra CP violation to the standard model in order to obtain an appreciable $n_B / s$ ratio. A simple possibility which has been invoked often is to add a second Higgs doublet to the theory, with CP violating relative phases. 

The standard way to obtain out-of-equilibrium baryon number violating processes immediately after the electroweak phase transition is$^{\cite{EWBGrev}}$ to assume that the transition is strongly first order and proceeds by the nucleation of bubbles (note that these are two assumptions, the second being stronger than the first!). 

Bubbles are out-of-equilibrium configurations. Outside of the bubble (in the false vacuum), the baryon number violating processes are unsuppressed, inside they are exponentially suppressed. In the bubble wall, the Higgs fields have a nontrivial profile, and hence (in models with additional CP violation in the Higgs sector) there is enhanced CP violation in the bubble wall. In order to obtain net baryon production, one may either use fermion scattering off bubble walls$^{\cite{CKN1}}$ (because of the CP violation in the scattering, this generates a lepton asymmetry outside the bubble which converts via sphalerons to a baryon asymmetry) or sphaleron processes in the bubble wall itself$^{\cite{TZ,CKN2}}$. It has been shown that, using optimistic parameters (in particular a large CP violating phase $\Delta \theta_{CP}$ in the Higgs sector) it is possible to generate the observed $n_B / s$ ratio. The resulting baryon to entropy ratio is of the order
\be \label{ewres}
{{n_B} \over s} \, \sim \, \alpha_W^2 (g^*)^{-1} \bigl( {{m_t} \over T} \bigr)^2 \Delta \theta_{CP} \, ,
\ee
where $\alpha_W$ refers to the electroweak interaction strength, $g^*$ is the number of spin degrees of freedom in thermal quilibrium at the time of the phase transition, and $m_t$ is the top quark mass. The dependence on the top quark mass enters because net baryogenesis only appears at the one-loop level.

However, analytical and numerical studies show that, for the large Higgs masses which are indicated by the current experimental bounds, the electroweak phase transition will unlikely be sufficiently strongly first order to proceed by bubble nucleation. In addition, there are some concerns as to whether it will proceed by bubble nucleation at all (see e.g. \cite{Gleiser}).

Once again, topological defects come to the rescue. In models which admit defects, such defects will inevitably be produced in a phase transition independent of its order. Moving topological defects can play the same
role in baryogenesis as nucleating bubbles. In the defect core, the electroweak symmetry is unbroken and hence sphaleron processes are unsuppressed$^{\cite{Perkins}}$. In the defect walls there is enhanced CP violation for the same reason as in bubble walls. Hence, at a fixed point in space, a nonvanishing baryon number will be produced when a topological defect passes by.

Defect-mediated electroweak baryogenesis has been worked out in detail in \cite{BDPT} (see \cite{BDT} for previous work) in the case of cosmic strings. The scenario is as follows: at a particular point $x$ in space, antibaryons are produced when the front side of the defect passes by. While $x$ is in the defect core, partial equilibration of $n_B$ takes place via sphaleron processes. As the back side of the defect passes by, the same number of baryons are produced as the number of antibaryons when the front side of the defect passes by. Thus, at the end a positive number of baryons are left behind.

As in the case of defect-mediated GUT baryogenesis, the strength of defect-mediated electroweak baryogenesis is suppressed by the ratio ${\rm SF}$ of the volume which is passed by defects divided by the total volume, i.e.
\be
{{n_B} \over s} \, \sim \, {\rm SF} {{n_B} \over s}|_0 \, ,
\ee
where $(n_B / s)|_0$ is the result of (\ref{ewres}) obtained in the bubble nucleation mechanism. 

A big caveat for defect-mediated electroweak baryogenesis is that the standard electroweak theory does not admit topological defects. However, in a theory with additional physics just above the electroweak scale it is possible to obtain defects (see e.g. \cite{TDB95} for some specific models). The closer the scale $\eta$ of the new physics is to the electroweak scale $\eta_{EW}$, the larger the volume in defects and the more efficient defect-mediated electroweak baryogenesis. Using the result of (\ref{defsep}) for the separation of defects, we obtain
\be
{\rm SF} \, \sim \, \lambda \bigl( {{\eta_{EW}} \over \eta} \bigr)^{3/2} \, .
\ee 

Obviously, the advantage of the defect-mediated baryongenesis scenario is that it does not depend on the order and on the detailed dynamics of the electroweak phase transition.  

\subsection{Summary}

As we have seen, topological defects may play an important role in cosmology.
Defects are inevitably produced during symmetry breaking phase transitions in the early Universe in all theories in which defects are topologically stable.
Theories giving rise to domain walls or local monopoles are ruled out by cosmological constraints. Those producing cosmic strings, global monopoles and textures are quite attractive.

If the scale of symmetry breaking at which the defects are produced is about $10^{16}$ GeV, then defects can act as the seeds for galaxy formation. Defect theories of structure formation predict a roughly scale-invariant spectrum of density perturbations, similar to inflation-based models. However, the phases in the density field are distributed in a non-Gaussian manner. Thus, the predictions of defect models can be distinguished from those of inflationary models. In addition, the predictions of different defect models can be distinguished from eachother.

Focusing on the cosmic string theory of structure formation, we have seen that the model gives rise to several distinctive signatures. The large-scale structure in the Universe is predicted to be dominated by either planar or filamentary structures (depending on whether there is small-scale structure on the strings), with a distinctive scale given by the comoving Hubble radius at $t_{eq}$. There are many ways topological measures with which one can quantify this prediction (see e.g. \cite{topology}). In the CMB, the distinctive signature for strings is predicted to be line discontinuities in the temperature maps.

Topological defects may also play a crucial role in baryogenesis. This applies
both to GUT and electroweak baryogenesis. The crucial point is that defects constitute out-of-equilibrium configurations, and may therefore be the sites of net baryon production.

\medskip
\centerline{\bf Acknowledgments}
\medskip

I wish to thank Professor Choonkyu Lee for inviting me to give these lectures  and all the organizers and participants for their wonderful hospitality in Korea. In particular, I am grateful to Dr. Jai-chan Hwang for interesting discussions, and to Bill Unruh for hosting me at the University of British Columbia where these lecture notes were written up.  
I also would like to all of my research collaborators, on whose work I have
freely drawn. Partial financial support for the preparation of this manuscript
has been provided at Brown by the US Department of Energy under Grant DE-FG0291ER40688,

}
\end{document}